\documentclass[aps,prb,twocolumn,showpacs,superscriptaddress,citeautoscript,longbibliography]{revtex4-2}
\usepackage{graphicx} % Include figure files
\usepackage{amsmath} % Useful mathematical tools
\usepackage{bm} % bold mathi
\usepackage{natbib}
\usepackage{soul}
\usepackage[normalem]{ulem}
\usepackage{natmove}
\usepackage[colorlinks=true, linkcolor=blue, citecolor=blue, urlcolor=blue, linktoc=page, bookmarks=false, pdfstartview={FitH}, pdfborder={0 0 0.0 [3 3]}]{hyperref} % HyperLink
\usepackage{xcolor} % For colored text
\usepackage{dcolumn} % Align table columns on decimal point
\newcolumntype{.}{D{.}{.}{2.1}}
\newcolumntype{-}{D{.}{.}{4.0}}
\usepackage{makecell}
\usepackage{multirow}
\usepackage{cleveref}
\Crefname{figure}{Fig.}{Figs}
\Crefname{table}{Table}{Tables}
\Crefname{section}{Sec.}{Sections}
\renewcommand{\today}{\number\day \space \ifcase \month \or January\or February\or March\or April\or May \or June\or July\or August\or September\or October\or November\or December\fi \space \number\year} % Date
\def\m1r{\multicolumn{1}{r}}

\begin{document}

% ========== TITLE ==========
\title{A DFT study of the structural and electronic properties of single and double acceptor dopants in MX$_2$ monolayers}
% Towards room temperature ferromagnetic semiconductor in p-doped MX2 monolayer
% ====================
% ========== AUTHORS AND AFFILIATIONS ==========
\author{Yuqiang Gao}
\affiliation{Department of Physics, School of Physics and Electronic Information, Anhui Normal University, Wuhu 241000,  PR China}
\affiliation{Faculty of Science and Technology and MESA$^+$ Institute for Nanotechnology, University of Twente, P.O.~Box~217, 7500~AE Enschede, The Netherlands}
\email[Email: ]{y.gao@ahnu.edu.cn}
\author{Paul J. Kelly\thanks{corresponding author}}
\email[Email: ]{P.J.Kelly@utwente.nl}
\affiliation{Faculty of Science and Technology and MESA$^+$ Institute for Nanotechnology, University of Twente, P.O.~Box~217, 7500~AE Enschede, The Netherlands}
% ====================
\date{\today}
% ========== ABSTRACT ==========
\begin{abstract}
Density functional theory calculations are used to systematically investigate the structural and electronic properties of MX$_2$ transition metal dichalcogenide monolayers with M = Cr, Mo, W and X = S, Se, Te that are doped with single (V, Nb, Ta) and double (Ti, Zr, Hf) acceptor dopants on the M site with local $D_{3h}$ symmetry in the dilute limit. Three impurity levels that arise from intervalley scattering are found above the valence band maxima (VBM): an orbitally doubly degenerate $e'$ level bound to the $K/K'$ VBM and a singly degenerate $a'_1$ level bound to the $\Gamma$-point VBM. 
Replacing S with Se or Te lowers the $\Gamma$ point VBM substantially with respect to the $K/K'$ VBM bringing the $a'_1$ level down with it. The relative positions of the impurity levels that determine the different structural and electronic properties of the impurities in $p$-doped MX$_2$ monolayers can thus be tuned by replacing S with Se or Te.

Single acceptors introduce a magnetic moment of 1$\, \mu_{\rm B}$ in all MX$_2$ monolayers. Out-of-plane magnetic anisotropy energies as large as 10 meV/dopant atom are found  thereby satisfying an essential condition for long-range ferromagnetic ordering in two dimensions. 
For double acceptors in MS$_2$ monolayers, both holes occupy the high-lying $a'_1$ level with opposite spins so there is no magnetic moment; 
in MSe$_2$ and MTe$_2$ monolayers the holes occupy the $e'$ level, a Jahn-Teller (JT) distortion wins the competition with exchange splitting resulting in the quenching of the magnetic moments. Even when the JT distortion is disallowed, magnetic double acceptors have a large in-plane magnetic anisotropy energy that is incompatible with long-range magnetic ordering in two dimensions.

The magnetic moments of pairs of single acceptors exhibit long-range ferromagnetic coupling except for MS$_2$ where the coupling is quenched for impurity pairs below a critical separation. For Se and Te compounds, the holes are accommodated in high-lying degenerate $e'$ levels which form triplets for all separations. However, for X=Te, a JT distortion lifts the degeneracy of the $e'$ levels leading to a reduction of the exchange interaction between impurity pairs. 
Deep, intrinsic, vacancy and antisite defects that localize the holes might stabilize the magnetization of $p$-doped MX$_2$ monolayers. Our systematic study of the $p$-doped MX$_2$ monolayers identifies 1H CrTe$_2$ and MoSe$_2$ as the most promising candidates for room temperature ferromagnetism.
We combine the exchange interaction estimated from the energy difference calculated for ferromagnetically and antiferromagnetically coupled pairs with Monte Carlo calculations to estimate the Curie temperatures $T_{\rm C}$ for vanadium doped CrTe$_2$ and MoSe$_2$ monolayers. Room temperature values of $T_{\rm C}$ are predicted for V dopant concentrations of 5\% and 9\%, respectively. In view of the instability of CrTe$_2$ in the 1H form, we suggest that the Cr$_x$Mo$_{1-x}$(Te$_y$Se$_{1-y})_2$ alloy system be studied.

A single $d$ electron or hole is uncorrelated. However, in the single impurity limit, the residual  self-interaction of this carrier in the local spin density approximation (LSDA) can be corrected by introducing a Hubbard $U$. Doing so leads to a large increase of the ordering temperatures calculated in the LSDA (reducing the doping concentration needed to achieve room temperature ordering) but at the expense of introducing an indeterminate parameter $U$.    

\end{abstract}
% ====================

 % Physics and Astronomy Classification Scheme
\maketitle

%%%%%%%%10%%%%%%%%20%%%%%%%%30%%%%%%%%40%%%%%%%%50%%%%%%%%60%%%%%%%%70%%%%%%%%80
% ========== INTRODUCTION ==========
\section{Introduction}
\label{sec:intro}
 
The observation of ferromagnetism in Mn-doped InAs \cite{Ohno:prl92} and GaAs \cite{Ohno:apl96} semiconductors and predictions that room temperature ordering might be attainable \cite{Dietl:sc00} stimulated attempts to realize such a dilute magnetic semiconductor (DMS). After twenty-five years and a very considerable research effort, the maximum ordering temperature has stagnated at values well below room temperature \cite{Dietl:natm10, Dietl:rmp14}.
Research proceeds largely empirically with the number of semiconducting materials being considered multiplying without it becoming clear what the fundamental limits of the ordering temperatures are that can be achieved in a certain material system. There are many reasons for the low ordering temperatures \cite{Jungwirth:rmp06, Sato:rmp10} but the main problem is that the open $d$ shell states of magnetic impurities like Mn are quite localized. While this favours the onsite exchange interaction which is the origin of the Hund's-rule spin alignment that makes the magnetic moment of the ion robust and insensitive to temperature, it leads to weaker exchange interactions between pairs of impurity ions that determine the Curie temperature $T_{\rm C}$, the ferromagnetic ordering temperature. Attempts to increase $T_{\rm C}$ by increasing the concentration of dopant atoms are thwarted by their clustering or by the formation of antisite defects, which are electron donors that reduce the concentration of holes. Transition metal ions such as Mn, Fe or Co induce ``deep levels'', tightly bound, partially occupied states in the fundamental gap of many semiconductors. At high dopant concentrations, these form deep impurity bands that dominate the (transport) properties of a material that, from the point of view of the electronic structure, is no longer a semiconductor but a completely new material. 
 
Recently we explored a different approach to realizing a high temperature magnetic semiconductor \cite{Gao:prb19a, *Gao:prb19b}. In a tight-binding picture, the transition metal (TM) $d$ bandwidth is $2zt$ where $z$ is the number of neighbouring TM atoms and $t$ is the nearest neighbour hopping matrix element. Because of the low coordination number in two-dimensional materials, the TM dichalcogenide (TMD) MoS$_2$ with $z=6$ has a substantially reduced $d$ bandwidth $W_d$. This makes it easier to satisfy the Stoner criterion for the occurrence of itinerant ferromagnetism, $D(E_F) I_{\rm xc} > 1$, where $I_{\rm xc}$ is the Stoner parameter and $D(E_F)$ is the density of states (DoS) at the Fermi level that is inversely proportional to the bandwidth. $I_{\rm xc}$ is essentially an atomic property and in the local spin density approximation (LSDA) \cite{vonBarth:jpc72, Gunnarsson:prb76} it contains contributions from both exchange and correlation \cite{Gunnarsson:jpf76, Andersen:physbc77a, Janak:prb77}.

\begin{figure}[t]
\includegraphics[width=8.6cm]{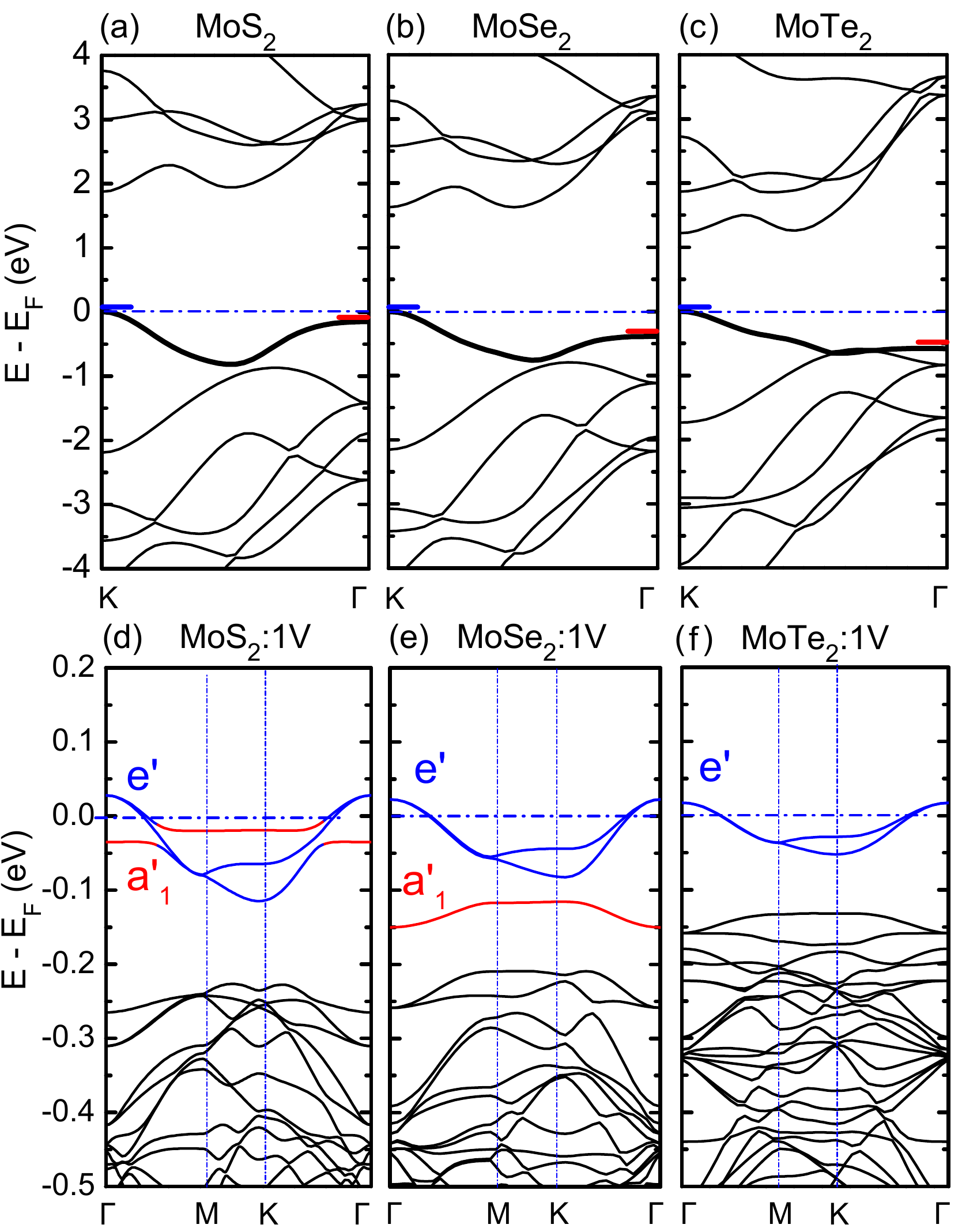}
\caption{Band structures without spin-orbit coupling of monolayers of undoped MoS$_2$ (a), MoSe$_2$ (b) and MoTe$_2$ (c) and with a single substitutional vanadium dopant atom on a Mo site in a 6$\times$6 supercell  (d-f). The zero of energy is the Fermi level. In (a-c), the blue and red horizontal lines represent, respectively, the $e'$ impurity levels bound to $\varepsilon(K/K')$, the $K/K'$ valence band maxima (VBM), and the $a'_1$ impurity level bound to $\varepsilon(\Gamma)$, the $\Gamma$ VBM. 
The energy difference $\varepsilon(K/K')-\varepsilon(\Gamma)=\Delta_{K\Gamma}$ increases in the sequence S$\rightarrow$Se$\rightarrow$Te. In (d-f) the blue and red lines are the corresponding bands formed from these impurity levels because of the finite size of the unit cell and the periodic boundary conditions. Note that different energy scales are used in the top and bottom panels.}
\label{fig:band}
\end{figure}

Mo has a 4$d^5$5$s^1$ electronic configuration and in MoS$_2$ it is nominally Mo$^{4+}$ with one up-spin and one down-spin $d$ electron. A monolayer of MoS$_2$ is a direct gap semiconductor with a very narrow Mo $d$ band containing these two electrons, the bold black band in \Cref{fig:band}(a), so it is nonmagnetic. This nonbonding band forms the top of the valence band in a gap formed by bonding and antibonding combinations of Mo-$d$ and S-$p$ states. To probe the enhanced DoS, we considered hole-doping MoS$_2$ by replacing some of the Mo atoms with group VB acceptor atoms \cite{Gao:prb19a, *Gao:prb19b}. 

% ========== Figure 1 ==========
% ====================

When a group VIB Mo atom is replaced by a group VB atom like V, Nb or Ta, then the dopant atom e.g. V$^{4+}$, has a single unpaired $d$ electron and therefore introduces a single hole into the narrow Mo 4$d$ band; substitution of a group IVB atom (Ti, Zr, Hf) will introduce two holes per dopant atom. In the single impurity limit, the asymptotic Coulomb potential of the ionized impurity leads to a series of hydrogenic states bound to the top of the valence band. At the $K(K')$ VBM, the highest such state is a doubly degenerate state with $e'$ symmetry in the local $D_{3h}$ symmetry that has mixed Mo $d_{x^2-y^2}$ and $d_{xy}$ character. At the slightly lower $\Gamma$-point VBM the highest state is an $a'_1$ symmetry state with Mo $d_{3z^2-r^2}$ character and a large effective mass \cite{Peelaers:prb12}. For finite concentrations of dopant atoms, these impurity states form impurity bands and at finite temperatures the holes in these impurity bands are thermally excited into the top of the narrow nonbonding Mo $d$ band. 

In \cite{Gao:prb19a, *Gao:prb19b} we found that pairs of V, Nb and Ta substitutional dopants in an MoS$_2$ monolayer coupled ferromagnetically for all dopant separations but the shortest. In the single impurity limit, a large out-of-plane magnetic anisotropy of $\sim 5\,$meV per dopant ion was found that allows circumvention of the Mermin-Wagner theorem according to which thermal fluctuations destroy ordering in two dimensions for isotropic Heisenberg exchange \cite{Mermin:prl66, Hohenberg:pr67}. Large single ion anisotropy (SIA) combined with isotropic exchange coupling leads to effective Ising-spin behaviour \cite{Leonel:jmmm06} and long-range magnetic ordering \cite{Onsager:pr44, Yang:pr52} at finite ordering temperatures that can be estimated using Monte-Carlo calculations \cite{Binder:zfpb81, Landau:09}. For V, Nb and Ta, we estimated values of $T_{\rm C}$ of about 100K for Nb and Ta and as high as 160K for V for doping concentrations of order 9\%. 
Very recent reports of long-range and/or room temperature ferromagnetism occurring in V-doped WS$_2$ \cite{Zhang:advs20} and WSe$_2$ monolayers \cite{Yun:advs20, Pham:am20}, in MoSe$_2$ and MoTe$_2$ \cite{Guguchia:sca18}, in V- and Ta-doped MoTe$_2$ \cite{Coelho:aem19, Yang:aem19} and in V-doped MoS$_2$ \cite{Hu:acsami19} (whereby the interaction with anion vacancies would appear to play an important role \cite{Hu:acsami19, Yun:am22}) provide the motivation to investigate these systems more systematically. 

In our study of MoS$_2$, we identified the formation of singlet states for dopant pairs closer than a critical separation ($\sim$ 9\AA) as the most important factor limiting the ordering temperature to below room temperature in $p$-doped MoS$_2$ monolayers \cite{Gao:prb19a, *Gao:prb19b}. For close dopant pairs, the $\pi$ interaction of $a'_1$ states with $d_{3z^2-r^2}$ character is stronger than the $\sigma$ interaction between $e'$ states with $d_{x^2-y^2}$ and $d_{xy}$ character so the highest-lying level occupied by the two holes is an antibonding $a^*$ state. It is orbitally nondegenerate so to satisfy the exclusion principle the holes are forced to have opposite spins i.e., the net magnetic moment is zero. Because of their $d_{3z^2-r^2}$ character, the $a'_1$ states are very localized in the $xy$ plane and the interaction between $a'_1$ states decays more rapidly than that between $e'$ states with $d_{x^2-y^2}$ and $d_{xy}$ character. For separations larger than the critical distance, the orbitally degenerate antibonding $e^*$ state can accomodate two holes with parallel spins corresponding to a ferromagnetic exchange interaction. 

For MoSe$_2$ and MoTe$_2$, the $\Gamma$-point VBM is much lower than for MoS$_2$, compare \Cref{fig:band}(a)-(c). We therefore expect that the $a'_1$ levels will lie correspondingly low in energy and be fully occupied so quenching of the ferromagnetic interaction will not occur. However, there is a downside. Because of its spatial localization, the exchange splitting of the $a'_1$ level is substantially larger than that of the $e'$ levels making it much less sensitive to electronic thermal excitation. This will become important when the exchange splitting acquires the physical meaning of a stabilization energy for  ferromagnetically coupled atoms. Because it is not a priori clear for which system the optimum interatomic exchange coupling and ordering temperature will occur, the purpose of this manuscript is to perform a systematic study of dopant acceptor levels in MX$_2$ monolayers with M = Cr, Mo, W; X = S, Se, Te in the single impurity limit. We will also evaluate the magnetic anisotropy in this same limit because of the Mermin-Wagner theorem. Intrinsic defects such as vacancies and antisite defects are commonly observed in various van der Waals materials. The interplay between substitutional M dopants and intrinsic defects could have significant influence on the Curie temperature so we will also consider some representative examples of these. 

The paper is organized as follows. After a brief summary of some computational details in \Cref{sec:CM}, the main results for the electronic and structural properties of $p$-doped MX$_2$ monolayers in the single impurity limit are reported in \Cref{sec:ResSIL}. The exchange interaction between single acceptor pairs and the consequences of these for room temperature ferromagnetism are examined in \Cref{sec:ResPI}. In \Cref{sec:disc}, we compare our results with those of other computational studies and consider experiments in the light of our results. \Cref{sec:sumconc} contains a brief summary and some conclusions. 

%\Cref{ssec:SEP}: Structural and electronic properties of p-doped MX$_2$ monolayers\\
%\Cref{sssec:FE}: Formation energy\\
%\Cref{sssec:BE}: Binding energy\\ 
%\Cref{sssec:SP}: Spin polarization \\ 
%\Cref{sssec:JT}: Jahn Teller distortions \\
%\Cref{sssec:SIA}: Single ion anisotropy \\
%
%\noindent
%\Cref{ssec:ID}: Interaction with intrinsic defects \\
%\Cref{ssec:ED}: Electron doping: Rhenium\\
%
%\noindent
%\Cref{sec:CM}: Computational Method\\
%\Cref{sec:results}: Results \\ 
%\Cref{sec:disc}: Discussion \\
%\Cref{sec:conc}: Summary & Conclusions

%%%%%%%%10%%%%%%%%20%%%%%%%%30%%%%%%%%40%%%%%%%%50%%%%%%%%60%%%%%%%%70%%%%%%%%80
%==========Computational Method=======
\section{Computational Method}
\label{sec:CM}

First-principles density functional theory (DFT) calculations were performed with a plane-wave basis and  the projector augmented wave (PAW) formalism \cite{Blochl:prb94b} as implemented in the {\sc vasp} code \cite{Kresse:prb93b, Kresse:prb96, Kresse:prb99}. An energy cut-off of 400 eV was used and atomic structures were fully relaxed until all forces were smaller than 0.01 eV/\AA. Monolayers of MX$_2$ were periodically repeated in the $c$ direction with a separation of more than 20~\AA\ of vacuum to avoid spurious interactions that arise from the use of (artificial) periodic boundary conditions. Because we only consider free-standing monolayers, there was no need to take van der Waals interactions into account. The local (spin) density approximation, L(S)DA, underestimates equilibrium atomic separations more than generalized gradient approximations (GGA) overestimate them. However, by comparison with experiment, the LDA describes the relative positions of the valence band maxima better than does the GGA \cite{Gao:prb19a, *Gao:prb19b}. Since this will play an important role in our study of shallow impurity levels, we used the LSDA as parameterized by Perdew and Zunger \cite{Perdew:prb81}. A 12$\times $12 supercell was used to reduce to an acceptable degree the interaction between impurities in neighboring supercells when modelling dopants in the single impurity limit. A 2$\times$2 k-point sampling was used for structure relaxation and 4$\times$4 for the calculations with spin-polarization and spin orbit coupling; a lower k-point sampling does not describe the magnetism sufficiently well. For intrinsic defects, a smaller 6$\times$6 supercell was found to be sufficient to describe the defect potential inside the supercell reliably because of the greater localization of the gap states. For structural relaxation, we used Gaussian smearing and a broadening of 0.01 eV \cite{Methfessel:prb89}. For more accurate total energies, the linear analytical tetrahedron method \cite{Jepsen:ssc71} with nonlinear corrections \cite{Blochl:prb94a} was used.

%%%%%%%%10%%%%%%%%20%%%%%%%%30%%%%%%%%40%%%%%%%%50%%%%%%%%60%%%%%%%%70%%%%%%%%80
%==========Results=========
\section{Results: Single Impurity Limit}
\label{sec:ResSIL}

We begin this section with a brief comparison of the calculated structural and electronic properties of pristine MX$_2$ monolayers in \Cref{ssec:MX2ML}. In \Cref{ssec:SEP}, we examine the structural and electronic properties of single and double acceptors in MX$_2$ monolayers. A number of intrinsic defects is considered in \Cref{ssec:ID} before briefly looking at the case of electron doping by the 5$d$ element rhenium in \Cref{ssec:ED}.

%\noindent
%Briefly state the contents of the Results section:
%\noindent
%\Cref{ssec:MX2ML}: MX$_2$ monolayers \\
%\Cref{ssec:SEP}: Structural and electronic properties of p-doped MX$_2$ monolayers\\
%\Cref{ssec:ID}: Interaction with intrinsic defects \\
%\Cref{ssec:ED}: Electron doping: Rhenium\\

%%%%%%%%10%%%%%%%%20%%%%%%%%30%%%%%%%%40--------50--------60--------70--------80
\subsection{MX$_2$ monolayers} 
\label{ssec:MX2ML}

% ========== Figure 2 ==========
\begin{figure}[b]
\includegraphics[width=8.6cm]{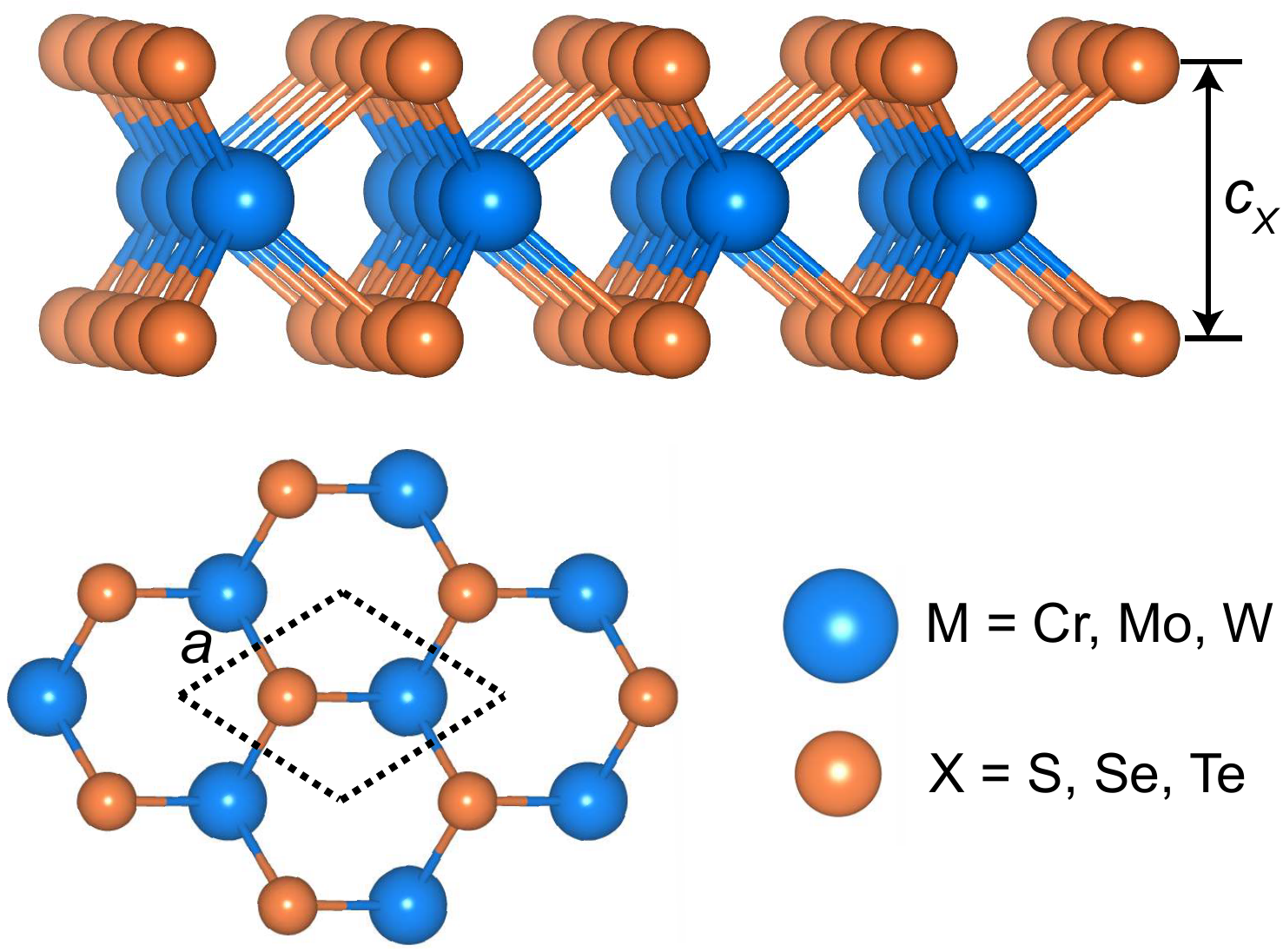}
\caption{\label{fig:structure}Side view and top view of the atomic structure of an MX$_2$ (M=Cr, Mo, W; X=S, Se, Te) monolayer. The unit cell is enclosed by the dotted lines.}
\end{figure}
% ====================

%==========Table1=========
\begin{table*}
\caption{Structural and electronic properties of 1H-MX$_2$ monolayers. Shown are the lattice constant $a$ and the separation of the two chalcogen X layers $c_{\rm X}$, both in \AA; the direct bandgap at the $K$ point $\Delta\varepsilon_g$, the energy difference $\Delta_{K\Gamma}=\varepsilon_\upsilon(K)-\varepsilon_\upsilon(\Gamma)$ between the VBM at $K$ and $\Gamma$ without (with) spin orbit coupling and the bandwidth $W$ of the ``nonbonding'' band forming the top of the valence band, all in eV. }
\begin{ruledtabular}
\begin{tabular}{llccccccccccccccc}
     & X & \multicolumn{5}{c}{S}
                        & \multicolumn{5}{c}{Se}
                                       & \multicolumn{5}{c}{Te} \\
\cline{2-2} \cline{3-7}                      \cline{8-12}                  \cline{13-17}
     & M  & $a$  & $c_{\rm X}$ 
                        & $\Delta\varepsilon_g$ 
                               & $\Delta_{K\Gamma}$ 
                                      & $W$  & $a$  & $c_{\rm X}$ 
                                                           & $\Delta\varepsilon_g$ 
                                                                  & $\Delta_{K\Gamma}$ 
                                                                         & $W$  & $a$  & $c_{\rm X}$ 
                                                                                              & $\Delta\varepsilon_g$ 
                                                                                                     & $\Delta_{K\Gamma}$ 
                                                                                                            & $W$ \\
\cline{1-7}                      \cline{8-12}                  \cline{13-17}
     & Cr & 2.97 & 2.92 & 1.07 & 0.22 (0.25) & 0.95 & 3.12 & 3.11 & 0.87 & 0.32 (0.36) & 0.86 & 3.37 & 3.38 & 0.62 & 0.45 (0.49) & 0.70 \\
Calc & Mo & 3.12 & 3.11 & 1.87 & 0.15 (0.22) & 1.06 & 3.25 & 3.31 & 1.62 & 0.38 (0.44) & 1.02 & 3.47 & 3.59 & 1.22 & 0.58 (0.66) & 0.86 \\
     & W  & 3.12 & 3.12 & 2.01 & 0.17 (0.35) & 1.30 & 3.25 & 3.31 & 1.72 & 0.43 (0.55) & 1.22 & -    & -    & -    & -    & - \\
\hline     
     & Cr &  -       &  -       &  -       &  -       &  - &  -       &  -       &  -       &  -       &  -       &  -       & -        &  -       & - & - \\
Expt & Mo & 3.16$^a$ & 3.17$^a$ & 1.90$^b$ & 0.14$^c$ & -  & 3.29$^a$ & 3.34$^a$ & 1.58$^d$ & 0.44$^e$ & 0.90$^e$ & 3.52$^f$ & 3.60$^f$ & 1.10$^g$ & - & - \\
     & W  & 3.15$^h$ & 3.14$^h$ & 2.01$^i$ & 0.24$^j$ & -  & 3.28$^h$ & 3.34$^h$ & 1.66$^k$ & 0.50$^e$ & 1.24$^e$ & -        & -        & -        & - & - \\     
\end{tabular}
\end{ruledtabular}
$^a$Ref.\onlinecite{Bronsema:zaac86}
$^b$Ref.\onlinecite{Mak:prl10}
$^c$Ref.\onlinecite{Jin:prl13}
$^d$Ref.\onlinecite{Zhang:natn14}
$^e$Ref.\onlinecite{Wilson:sca17}
$^f$Ref.\onlinecite{Knop:cjc61}
$^g$Ref.\onlinecite{Ruppert:nanol14}
$^h$Ref.\onlinecite{Schutte:cjc87}
$^i$Ref.\onlinecite{Tongay:nanol14}
$^j$Ref.\onlinecite{Kastl:tdm18}
$^k$Ref.\onlinecite{Hsu:natc17}
\label{TableI}
\end{table*}
%==========Table1=========

Under ambient conditions, MoX$_2$ and WX$_2$ with X = S, Se, Te have a so-called ``H'' structure in which a hexagonal plane of M atoms is sandwiched between layers of X atoms such that each M atom sits at the centre of a trigonal prism with local $D_{3h}$ symmetry, \Cref{fig:structure}. Bulk structures are composed of van der Waals bound stacks of such layers in the $z$ direction perpendicular to the basal planes with one (1H) or two (2H) layers in the periodic unit cell \cite{Manzeli:nrm17}.
CrSe$_2$ is reported to be have been prepared in a metastable ``T'' structure \cite{vanBruggen:physbc80} where one of the X-atom planes is rotated about the $z$ axis by 180$^\circ$ so that each Cr atom sits at the centre of a regular octahedron of X atoms.  
1T CrS$_2$ and CrSe$_2$ are claimed to be metastable \cite{Fang:jpcm97} and a mixed 1H, 1T, and 1T$'$ phase of CrS$_2$ was reported to be formed by chemical vapor decomposition \cite{Habib:nanos19}. 
To realize a ferromagnetic semiconductor, we will focus on doping the semiconducting 1H phase MX$_2$ monolayers and will include CrX$_2$ for the purposes of comparison. Experimentally, a monolayer of WTe$_2$ is found to be semi-metallic with a $T_d$ polytype structure \cite{Wilson:ap69} and will not be considered further in this manuscript. 

The equilibrium structural and electronic parameters for 1H-MX$_2$ monolayers as calculated in the LDA \cite{Perdew:prb81} are listed in \Cref{TableI} and, where possible, compared to experiment. The lattice constant $a$ is seen to increase as X goes from S$\, \rightarrow \,$Se$\,\rightarrow \,$Te. This leads to a reduction in the overlap of M $d$ orbitals and consequently of the bandwidth $W$ of the ``nonbonding'' $d$ band that forms the topmost valence band in \Cref{fig:band}(a)-(c). The direct band gap $\Delta \varepsilon_g$ at the $K$ point decreases from MS$_2$ to MTe$_2$, while the energy difference $\Delta_{K\Gamma}=\varepsilon_\upsilon(K)-\varepsilon_\upsilon(\Gamma)$ between the $K/K'$ and $\Gamma$ VBM   increases in the same sequence both with and without SOC. The experimental $\Delta_{K\Gamma}$ listed in \Cref{TableI} are obtained from ARPES measurements and include SOC. 

Even though the lattice constant increases in the sequence CrX$_2$$\rightarrow$MoX$_2$$\rightarrow$WX$_2$, as M is changed from Cr to W the bandwidth $W$ increases. The overlap of the M $d$ orbitals that determines $W$ is related not only to the M-M bond length but is also determined by the spatial extent of the outermost $d$ orbitals. As we progress from $3d \rightarrow 4d \rightarrow 5d$, the outermost $d$ orbitals must be orthogonal to the more strongly bound lower-lying ones, are therefore excluded from the region of space closest to the nucleus and end up having a larger spatial extent. The band gap $\Delta\varepsilon_g$ also increases with increasing atomic number of M because the greater spatial extent of the $d$ orbitals leads to a stronger bonding-antibonding interaction with the X$\,p$ orbitals pushing the uppermost four empty $d$ states higher in energy. $\Delta_{K\Gamma}$ exhibits a similar trend except for the CrS$_2$ monolayer. In the next section, we will see that the properties of  dopants are largely determined by those of the host material suggesting directions to explore to increase the ferromagnetic ordering temperatures.

%%%%%%%%10%%%%%%%%20%%%%%%%%30%%%%%%%%40--------50--------60--------70--------80
\subsection{Structural and electronic properties of p-doped MX$_2$ monolayers in the single impurity limit} 
\label{ssec:SEP}

We begin this section by calculating the formation energy of single and double acceptors in MX$_2$ monolayers, \Cref{sssec:FE}. 
The impurity binding energies calculated relative to the VBM are presented in \Cref{sssec:BE} while the magnetic moments and exchange splittings calculated for each dopant are presented in \Cref{sssec:SP}. 
Before considering JT distortions for single and double acceptors in \Cref{sssec:JTsa,sssec:JTda}, we briefly consider the effect of symmetry-preserving relaxation in \Cref{sssec:SR}. 
Single ion anisotropy energies are presented in \Cref{sssec:SIA}.
 
%\noindent 
%Briefly state the contents of this subsection:\\
%\Cref{sssec:FE}: Formation energy\\
%\Cref{sssec:BE}: Binding energy\\ 
%\Cref{sssec:SP}: Spin polarization \\ 
%\Cref{sssec:JT}: Jahn Teller distortions \\
%\Cref{sssec:SIA}: Single ion anisotropy \\

%-------10--------20--------30--------40--------50--------60--------70--------80
\subsubsection{Formation energy} 
\label{sssec:FE}

The formation energy of an M$'$ dopant atom on an M atom host site, defined as 
\begin{equation}
\label{eq:form}
E_{\rm form}[{\rm M'_M}] = E_{\rm tot}[{\rm MX_2\!:\!M'}] - E_{\rm tot}[{\rm MX_2}] 
                                              + \mu_{\rm M} - \mu_{\rm M'}        \\
\end{equation}
determines how easily doping is achieved. $E_{\rm tot}[{\rm MX_2\!:\!M'}]$ is the total energy of a relaxed supercell of an MX$_2$ monolayer with a single M atom replaced with an M$'$ dopant atom. $E_{\rm tot}[\rm MX_2]$ is the total energy of the same size supercell but for a pristine MX$_2$ monolayer. $\mu_{\rm M}$ and $\mu_{\rm M'}$ are the chemical potentials of M and M$'$ atoms in their bulk bcc, respectively bcc or hexagonal, metallic phases. A negative formation energy in \Cref{TableII} indicates that substitutional doping of a group IVB or VB element in a monolayer of MX$_2$ is energetically favorable. The positive formation energies for some V-doped MX$_2$ monolayers are quite small relative to the magnitude of the heat of formation of MX$_2$ indicating that V doping is experimentally feasible. Globally, the formation energy increases in size from V to Ta, reaches a maximum size for Ti and is smaller for Zr and Hf. The formation energies of double acceptors are larger than those of single acceptors.

Substituting chalcogen $X$ atoms with $M$ atoms or TM impurities in $MX_2$ would create antisite defects, $M_X$, as well as $TM_X$. As indicated in \cite{Onofrio:jap17} and \Cref{ssec:ID}, this will be energetically unfavorable compared to the substitution of $M$ atoms. According to  \cite{Karthikeyan:nanol19}, TM impurities occupying the chalcogen site, $TM_X$, can be energetically favorable in postsynthesis impurity introduction. However, in the present work, we focus on substituting M atoms with shallow impurities $M'$ by impurity co-deposition whereby holes can be introduced into the valence band maximum of $MX_2$ instead of into trapped (deep) impurity levels. 

% ========== Table 2 ==========%
\begin{table} [b]
\caption{Calculated formation energies in eV of various substitutional dopants M$'$ on M sites (M$'_{\rm M}$) in MX$_2$ monolayers with relaxation. }
\begin{ruledtabular}
\begin{tabular}{lrrrrrrrrr}
   & M  & Cr  &   Mo  &   W   &   Cr  &   Mo  &   W   &   Cr  &   Mo  \\                                                  
   \cline{3-5} \cline{6-8} \cline{9-10}
M$'$ &  X   &  \multicolumn{3}{c}{S}
                            & \multicolumn{3}{c}{Se}
                                                    & \multicolumn{2}{c}{Te} \\
\hline
V   & & -0.26 &  0.25 &  0.24 & -0.37 &  0.32 & -0.07 & -0.45 &  0.22 \\
Nb  & & -0.25 & -0.15 & -0.10 & -0.55 & -0.14 & -0.49 & -0.62 &  0.03 \\
Ta  & & -0.36 & -0.23 & -0.02 & -0.43 & -0.10 & -0.39 & -0.57 &  0.07 \\
Ti  & & -0.32 & -0.58 & -0.64 & -0.74 & -0.57 & -0.72 & -0.62 & -0.33 \\
Zr  & & -0.25 & -0.25 & -0.21 & -0.46 & -0.27 & -0.55 & -0.56 & -0.21 \\
Hf  & & -0.42 & -0.32 & -0.39 & -0.52 & -0.31 & -0.84 & -0.59 & -0.42  
\end{tabular}
\end{ruledtabular}
\label{TableII}
\end{table}
% ========== Table 2 ==========% 

%-------10--------20--------30--------40--------50--------60--------70--------80
\subsubsection{Binding energy} 
\label{sssec:BE}

In effective mass theory (EMT), attractive and repulsive Coulomb potentials lead to the formation of Rydberg series of (shallow impurity) levels $\varepsilon_n$ with $n=1,2,3 ...$ bound to the conduction band minima and valence band maxima, respectively \cite{Pantelides:rmp78, Lannoo:81, Altarelli:82}. On substituting an M atom (M = Cr, Mo, W) in an MX$_2$ monolayer with a single (V, Nb, Ta) or a double (Ti, Zr, Hf) acceptor in the single impurity limit, the repulsive impurity potential and intervalley scattering leads to the formation of an orbitally twofold degenerate $e'$ level and a singly degenerate $a'_1$ level bound to the $K/K'$ and $\Gamma$ VBM, respectively \cite{Gao:prb19a, *Gao:prb19b}. The higher ($n \geq 2$) impurity levels are too shallow to show up above the VBM in the 12$\times$12 supercell size we standardly use \cite{Gao:prb19b} and therefore only the $e'$ and $a'_1$ impurity levels will be considered here. The impurity binding energy, defined as the position of the level with respect to the VBM, determines the degree of localization of the impurity wavefunction.

% ========== Figure 3 ==========
\begin{figure*}[tbp]
\includegraphics[scale = 0.57]{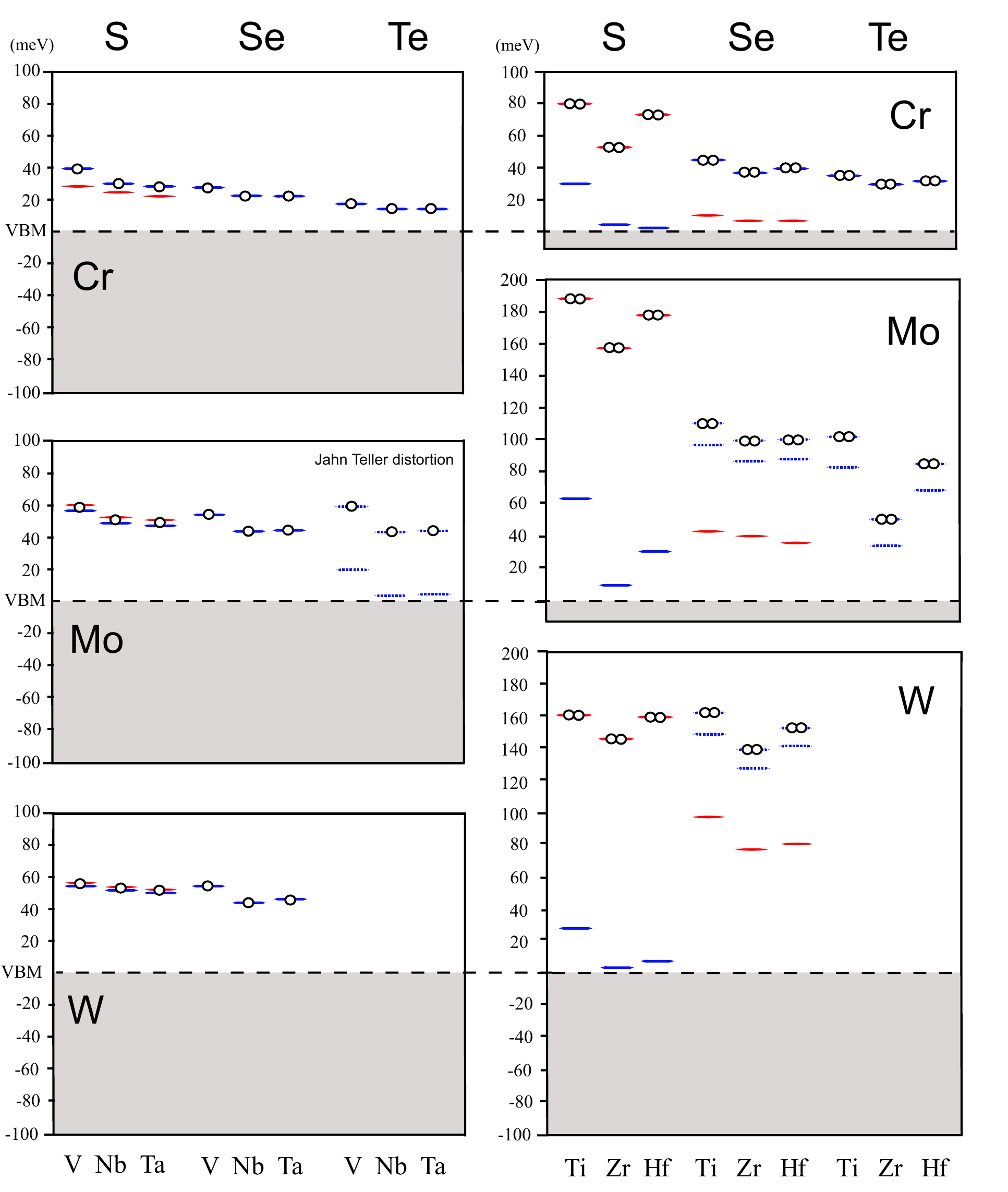}
\caption{\label{fig:impuritylevel} 
Impurity levels for single (V, Nb, Ta) and double (Ti, Zr, Hf) acceptors in MX$_2$ monolayers including atomic relaxation but without spin polarization. All energies are shown in meV with respect to the top of the valence band. The short horizontal red and blue lines represent $a'_1$ and degenerate $e'$ levels, respectively. The dotted blue lines indicate the Jahn-Teller split $e'$ levels. Holes represented by open circles occupy the highest levels in each system. Note that the smallest band gap is large on the energy scale of the figure. }
\end{figure*}
% ====================

To establish the position of impurity levels with respect to the VBM, we first need to be able to identify the VBM in a doped system, something which is not trivial in a supercell calculation because of the interaction of Coulomb bound states with their periodic images and the overlapping of the resulting impurity bands with the VBM \cite{Gao:prb19b}. First, because the binding energies of both single and double acceptors in MX$_2$ monolayers, shown in \Cref{fig:impuritylevel}, are much smaller than the sizeable bandgaps seen in \Cref{TableI}, there is no overlap between the impurity bands and the conduction band minima (CBM). 
Second, because a repulsive Coulomb potential does not affect the conduction band states, in particular the states forming the CBM, we can determine the bulk VBM in a doped system from its position relative to the CBM in the pristine case ($\varepsilon_{\rm VBM}$= $\varepsilon_{\rm CBM}- \Delta\varepsilon_g$). Alternatively, we can align the VBM to a core level of an atom far from the dopant site, e.g., the Mo 4$s$ level, extrapolating to infinite separation and relating that to the corresponding energy separation of the core level and VBM in a pristine MX$_2$ monolayer. The two procedures yield consistent results \cite{Gao:prb19b}. Because of its greater convenience, we use the first procedure to extract the position of the VBM here. The position of single (V, Nb, Ta) and double (Ti, Zr, Hf) acceptor impurity levels with respect to the VBM in doped MX$_2$ monolayers determined in this way are shown in \Cref{fig:impuritylevel}. Because of the finite dispersion of the impurity bands in a supercell calculation, what is shown is the appropriate weighted average over the Brillouin zone of the impurity band.

For monolayers of MoS$_2$ \cite{Gao:prb19a, Gao:prb19b}, the $a'_1$ and $e'$ levels are almost degenerate and the binding energy of single acceptor levels decreases from V to Ta as shown on the left-hand side of \Cref{fig:impuritylevel} (lhs, middle). The energy difference $\Delta_{K\Gamma}$ between the $K/K'$ and $\Gamma$ VBM in defect-free MX$_2$ monolayers determines the relative position of the impurity levels bound to these VBM. As shown in \Cref{fig:band} and tabulated in \Cref{TableI}, $\Delta_{K\Gamma}$ increases as X goes from S to Te, so the $a'_1$ levels (red lines) will drop fast with respect to the $e'$ levels (blue lines) going from MoS$_2$ to MoTe$_2$, which makes the $a'_1$ levels in MoSe$_2$ and MoTe$_2$ merge with the VBM in the $12 \times 12$ supercell we use, \Cref{fig:impuritylevel}. The $e'$ levels drop slightly from MoS$_2$ to MoTe$_2$. In MoTe$_2$, a Jahn-Teller distortion splits the degenerate $e'$ levels; we return to this in \Cref{sssec:JTsa}. 

In single-acceptor doped CrX$_2$ monolayers (top left), the impurity binding energies exhibit trends similar to those found in MoX$_2$. 
However the binding energies are lower and for CrS$_2$ the $e'$ level is bound more strongly than the $a'_1$ level so the hole occupies the former. 
In WX$_2$ (bottom left) the impurity binding energies are almost the same as in MoX$_2$. 

For the double acceptor levels shown on the right-hand side (rhs) of \Cref{fig:impuritylevel}, the main trends for the $a'_1$ and $e'$ levels in MX$_2$ parallel those found for single acceptors. 
The main difference is that the impurity binding energies of double acceptors are larger than those of single acceptors; the stronger repulsive potentials of the M$'^{2-}$ ions (attractive for holes) affect the localized out-of-plane $a'_1$ levels more than the in-plane $e'$ states. 
The $a'_1$ levels with $d_{3z^2-r^2}$ character lie well above the $e'$ levels and accommodate both holes in Ti, Zr and Hf-doped MS$_2$. 
In MSe$_2$ and MTe$_2$ they follow the $\Gamma$-point VBM down in energy, losing their holes to the orbitally degenerate $e'$ levels that, being partly filled, are susceptible to Jahn-Teller distortion. 
The competition between Jahn-Teller distortion and spin polarization will be discussed in \Cref{sssec:JTda} for double acceptor doped MSe$_2$ and MTe$_2$.  

%-------10--------20--------30--------40--------50--------60--------70--------80
\subsubsection{Spin polarization} 
\label{sssec:SP}

The results of calculating the magnetic moments of MX$_2$ monolayers with different dopants in 12$\times$12 supercells including full structural relaxation are given in \Cref{TableIII}. In the single impurity limit, the magnetic moments are expected to be an integer number of Bohr magnetons, $\mu_{\rm B}$. For all three single acceptors V, Nb, and Ta in MoS$_2$, this is true and the polarization is complete. The same is true for MoSe$_2$ and MoTe$_2$, \Cref{fig:impuritylevel} (lhs, middle). 

CrX$_2$ monolayers doped with single acceptors also have a magnetic moment of 1$\mu_{\rm B}$ for all three single acceptors, (lhs, top). Even though the defect levels of Nb and Ta acceptors in WX$_2$ (lhs, bottom) are as deep as in MoX$_2$ and deeper than in CrX$_2$, see \Cref{fig:impuritylevel} (lhs), their magnetic moments are less than 1$\mu_{\rm B}$. This incomplete polarization can be attributed to the use of a finite, 12$\times$12 supercell. If a supercell is not sufficiently large, then the interaction between dopants in neighbouring cells leads to the formation of bands by the discrete impurity levels. Unless the exchange splitting is larger than the impurity bandwidth, this will result in noninteger values of $n_\uparrow$ and $n_\downarrow$ and consequently in their difference $m= n_\uparrow - n_\downarrow$ being noninteger \cite{footnote2}. Even when dopants in WX$_2$ are fully polarized in the single impurity limit, as is the case for V and Ti, their exchange splittings may be smaller than those in CrX$_2$ and MoX$_2$ (see \Cref{TableIV}). As we discussed above, the greater spatial extent of the W 5$d$ orbitals leads to wider impurity bands for a given size of supercell.  Similarly, the greater interaction between impurity states and host W $d$ orbitals renders the impurity states more delocalized. For WX$_2$ doped with Nb and Ta, a larger supercell is needed to reach the single impurity limit; in the present case, a magnetic moment of 1$\mu_{\rm B}$ is found with a 15$\times$15 supercell. 

% ========== Table III ==========%
\begin{table} [tbp]
\caption{Calculated magnetic moments in $\mu_{\rm B}$ for monolayers of MX$_2$ ($12 \times 12$ supercell) doped with various transition metal atoms M$'$ substituting a single M. The structures are fully relaxed with spin polarization. Jahn-Teller distortion is included for single- but not for double-acceptors; the competition between spin-polarization and Jahn-Teller splitting will be found to quench the magnetic moments in double-acceptor doped MoSe$_2$, WSe$_2$, and MoTe$_2$ monolayers (\it italics).}
\begin{ruledtabular}
\begin{tabular}{llcccccccc}
  & M &  Cr  &  Mo  &  W   &  Cr  &  Mo  &  W   &  Cr  &   Mo  \\                                                  
      \cline{3-5} \cline {6-8} \cline {9-10}
M$'$&X&  \multicolumn{3}{c}{S}
                         & \multicolumn{3}{c}{Se}
                                              & \multicolumn{2}{c}{Te} \\
\hline
V   & & 1.00 & 1.00 & 1.00 & 1.00 & 1.00 & 1.00 & 1.00 & 1.00 \\
Nb  & & 1.00 & 1.00 & 0.91 & 1.00 & 1.00 & 0.71 & 1.00 & 1.00 \\
Ta  & & 1.00 & 1.00 & 0.90 & 1.00 & 1.00 & 0.70 & 1.00 & 1.00 \\
\hline
Ti  & & 0.00 & 0.00 & 0.00 & 2.00 & (\it 2.00) & (\it 2.00) & 2.00 & (\it 2.00) \\
Zr  & & 0.00 & 0.00 & 0.00 & 2.00 & (\it 2.00) & (\it 2.00) & 2.00 & (\it 2.00) \\
Hf  & & 0.00 & 0.00 & 0.00 & 2.00 & (\it 2.00) & (\it 2.00) & 2.00 & (\it 2.00) \\
\end{tabular}
\end{ruledtabular}
\label{TableIII}
\end{table}

% ========== Table IV ==========%
\begin{table} [b]
\caption{Exchange splittings $\Delta$ (in meV) calculated with a 12$\times$12 supercell for monolayers of MX$_2$ in which a single M atom is replaced by V or Ti. Jahn-Teller distortion is included for single but not for double acceptors. 
}
\begin{ruledtabular}
\begin{tabular}{lcrrrrrrrr}
   &   M    &   Cr &  Mo  &   W  &  Cr  &  Mo  &   W  &  Cr  &  Mo  \\                                                  
    \cline{3-5} \cline {6-8} \cline {9-10}
M$'$&  X    &  \multicolumn{3}{c}{S}
                                 & \multicolumn{3}{c}{Se}
                                                      & \multicolumn{2}{c}{Te} \\
\hline
 V & $a'_1$ & 34.5 & 96.1 & 91.2 &  -   & - & - & -    & -    \\
   & $e'$   & 21.5 & 15.2 &  9.3 & 24.2 & 26.7 & 12.0 & 24.1 & 44.0 \\
 \hline
Ti & $a'_1$ &  0.0 &  0.0 &  0.0 & 71.3 & 86.5 & 93.8 &  -   &  -   \\
   & $e'$   &  0.0 &  0.0 &  0.0 & 73.3 & 84.8 & 64.4 & 71.4 & 90.2 \\

\end{tabular}
\end{ruledtabular}
\label{TableIV}
\end{table}
The double acceptor impurities are all fully polarized with magnetic moments of $2\mu_{\rm B}$ except for the disulphide MS$_2$ monolayers with M = Cr, Mo, W. In these systems, the $a'_1$ levels lies well above the $e'$ levels, as seen in \Cref{fig:impuritylevel} (rhs), leaving two holes to occupy the singly degenerate $a'_1$ level with opposite spins and no magnetic moment. In the MSe$_2$ and MTe$_2$ monolayers, the holes are accommodated in doubly degenerate $e'$ levels and can have the same spin. However, the half-filled orbitally degenerate state is susceptible to a Jahn-Teller (JT) distortion and there is a competition between exchange interaction and JT distortion. If the JT distortion is strong enough, the magnetic moment will be quenched. This will be discussed in \Cref{sssec:JTsa}. 

The exchange splittings $\Delta$ calculated for monolayers of MX$_2$ doped with V and Ti are listed in \Cref{TableIV}. The size of $\Delta$ depends on the spatial localization of the corresponding states which in turn depends on the binding energies and orbital characters of the $a'_1$ and $e'$ levels. 
For V doped MS$_2$ monolayers, the exchange splitting of the $a'_1$ level is larger than that of the $e'$ level because of its greater spatial localization. 
In V doped MSe$_2$ and MoTe$_2$ monolayers, the $a'_1$ levels overlap with the top of the valence band, \Cref{fig:impuritylevel}, making it difficult to identify the corresponding exchange splittings which are not given in \Cref{TableIV}. 
In the sequence MS$_2$$\rightarrow$MSe$_2$$\rightarrow$MTe$_2$, the exchange splittings of the $e'$ levels are seen to increase (but not for CrSe$_2$:Ti $\rightarrow$ CrTe$_2$:Ti).

The exchange splittings of the Ti double acceptor levels are much larger than those of the V single acceptor because the impurity levels are much deeper and the hole densities much higher. As shown in \Cref{fig:impuritylevel}, the $e'$ impurity levels of double acceptors in MSe$_2$ are deeper than the $a'_1$ levels making their wave functions more localized and their hole densities larger. Therefore, the exchange splittings of the $e'$ levels are close to (or larger than) those of the $a'_1$ levels, \Cref{TableIV}. In CrTe$_2$ and MoTe$_2$ monolayers, the $a'_1$ levels are too shallow to be identified (\Cref{fig:impuritylevel}) and the corresponding exchange splittings are not shown in the Table.    

%-------10--------20--------30--------40--------50--------60--------70--------80
\subsubsection{Symmetric Relaxation}
\label{sssec:SR}

When a group VIB metal atom M in an MX$_2$ monolayer is substituted with a group IVB or VB element, the difference between the atomic radii of the dopant M$'$ and host M atoms leads to a change in the bond length of the dopant with the neighbouring chalcogen atoms. 
Symmetric structure relaxations that maintain the local $D_{3h}$ symmetry of the dopant atom on the high symmetry host site are found for all of the doped systems. 
For example, when a Mo atom in MoS$_2$ is replaced with a smaller V atom the surrounding S atoms move towards the V impurity causing the neighbouring Mo atoms to move slightly away \cite{Gao:prb19b}. 
For the unrelaxed impurity, the $a'_1$ and $e'$ levels are essentially degenerate. 
The relaxation shifts the $e'$ level slightly below the $a'_1$ level as seen in \Cref{fig:impuritylevel} (and in Fig.~10 in \cite{Gao:prb19b}). 
The total energy gain of $\sim$$150$~meV from the symmetric relaxation about an MoS$_2$:V impurity found in \cite{Gao:prb19b}, is much larger than the gain of single-particle energy associated with the shift of the $e'$ level indicating that a ``defect molecule'' picture \cite{Lannoo:81} is greatly oversimplified. 
It is an order of magnitude larger than the energy gained from spin polarization. 
Similar behaviour is found for Nb and Ta impurities though because of their larger atomic radii the chalcogen atoms move away from them. 
The same pattern of symmetric relaxation appears in MoSe$_2$ but in the next subsection we will see that single acceptors in MoTe$_2$ exhibit an additional, non-symmetric, Jahn-Teller, distortion.  

The atomic volumes of the double acceptors Ti, Zr and Hf are larger than those of Cr, Mo and W and we find outward symmetric relaxations of the neighbouring S ligands and transition metals in MS$_2$ monolayers which are consistent with the ionic radii; these shift the $e'$ levels much further below the $a'_1$ level than happens for single acceptors, \Cref{fig:impuritylevel}. 
When spin polarization is included simultaneously with relaxation for double acceptor-doped MSe$_2$ and MTe$_2$ monolayers, then the relaxation is found to be symmetric. 
Because of the large values of $\Delta_{K\Gamma}$ for MSe$_2$ and MTe$_2$ (\Cref{TableI}),  the $e'$ levels bound to the $K/K'$ valley still lie above the $a'_1$ levels bound to the $\Gamma$ VBM after the symmetric relaxation, \Cref{fig:impuritylevel}. 
The driving force for the symmetric relaxation is from the atomic interaction between the dopant and the atoms surrounding it and just as we found for single acceptors, the electronic energy gain is much larger than that associated with the (shallow) impurity levels in the gap. However, in some systems like V doped MoTe$_2$ with partially occupied degenerate impurity levels, the energy gain from the splitting of degenerate impurity levels, the Jahn-Teller distortion, leads to additional atomic relaxation that we consider next. 

%-------10--------20--------30--------40--------50--------60--------70--------80
\subsubsection{Jahn-Teller distortions: single acceptors}
\label{sssec:JTsa}

% ========== Figure 4 ==========
\begin{figure}[t]
\includegraphics[width=7.2cm]{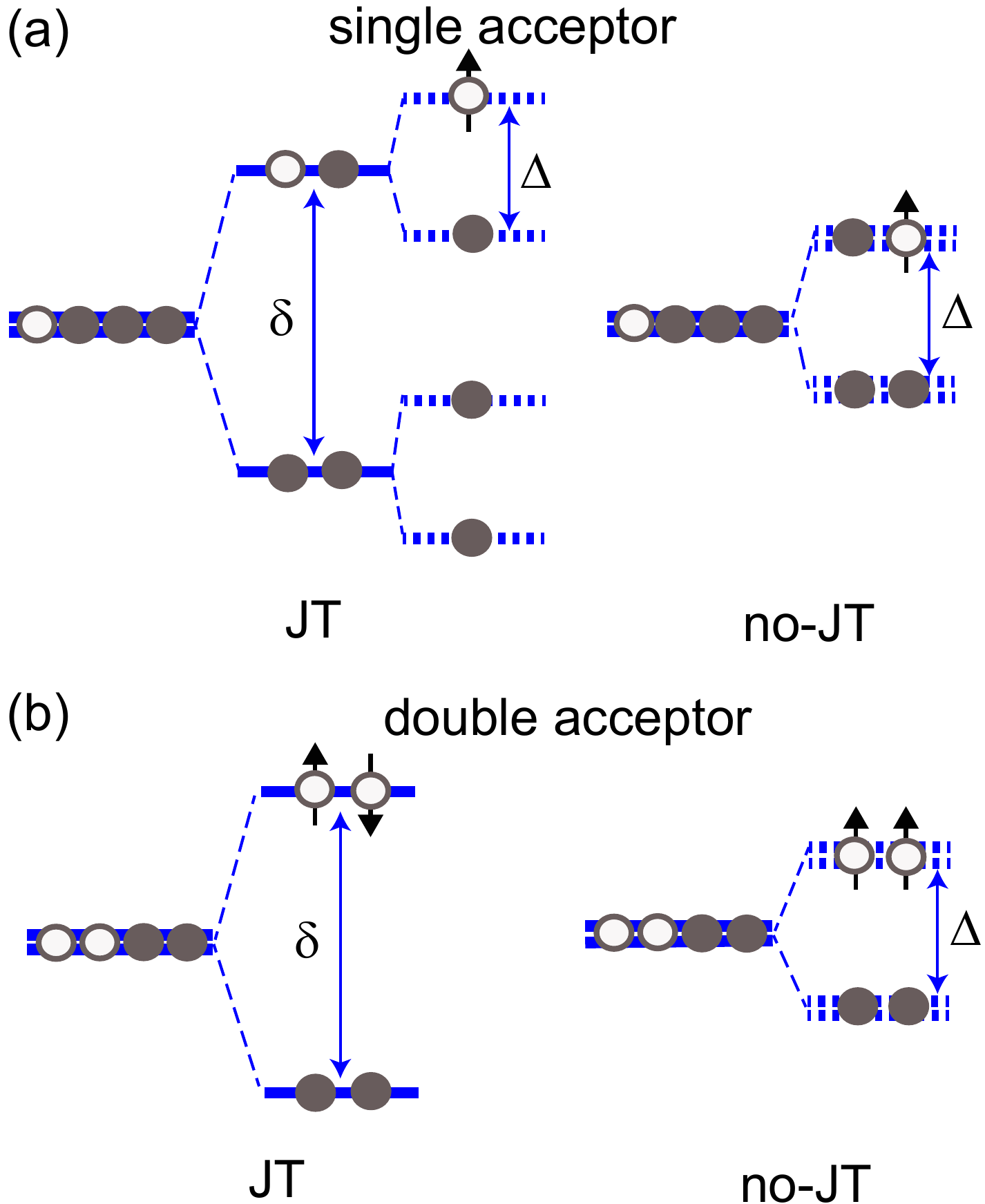}
\caption{\label{fig:JT} Schematic of the energy gain with and without Jahn-Teller distortion for (a) single and (b) double acceptors. The solid and dashed blue lines represent spin-degenerate and spin-resolved energy levels, respectively. $\delta$ denotes the energy splitting of the degenerate $e'$ levels caused by JT distortion and  $\Delta$ is the exchange splitting. Solid and open circles denote electrons and holes, respectively. The completely filled (with electrons; empty of holes) $a'_1$ level is not shown.}
\end{figure}
% ====================
% ========== Figure 5 ==========
\begin{figure}[b]
\includegraphics[width=5.2cm]{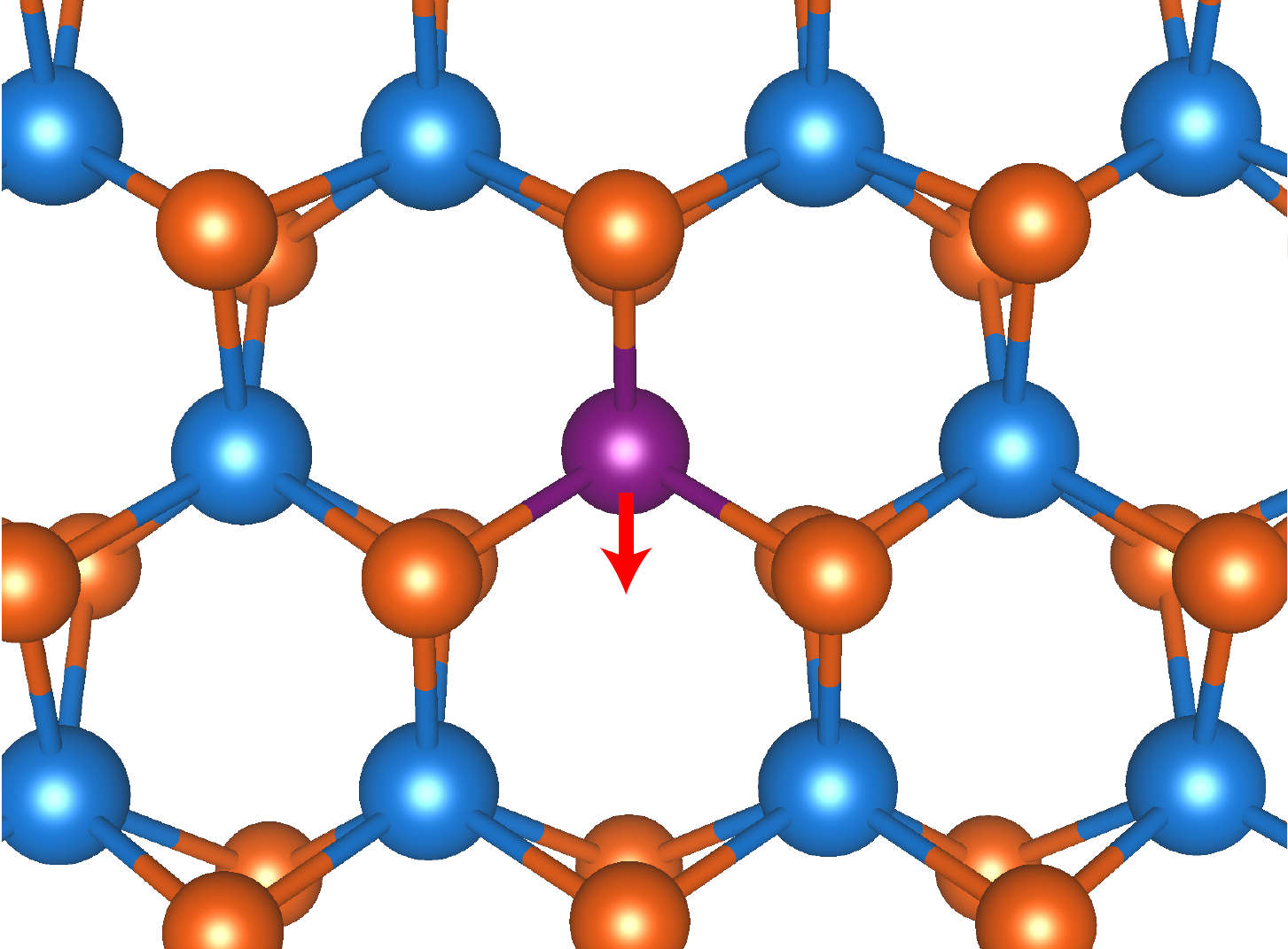}
\caption{\label{fig:JT_structure}Top view of the dopant-induced Jahn-Teller distortion in an MX$_2$ monolayer. The purple ball represents the dopant atom and the red arrow indicates its displacement for single acceptors; for double acceptors, the displacement is in the opposite direction. Because of the threefold symmetry of the unrelaxed structure, the symmetry-lowering dopant displacement can be in any one of three equivalent directions.} 
\end{figure}
% ====================

% ========== Figure 6 ==========
\begin{figure*}[t]
\includegraphics[width=17.6cm]{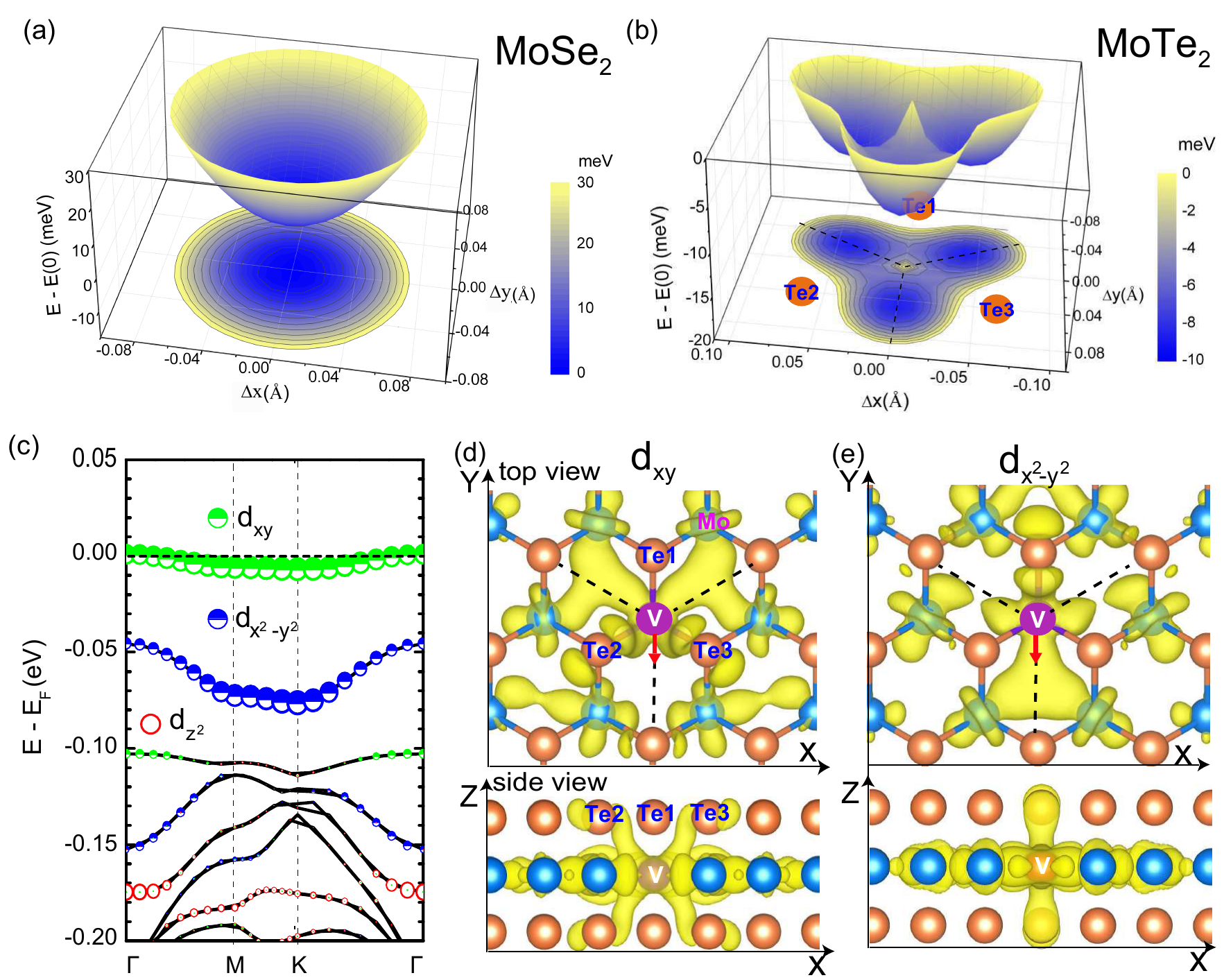}
\caption{\label{fig:PES} Total energy on displacing a substitutional vanadium dopant, ${\rm V_{Mo}}$, in the central, $xy$ metal plane away from the Mo lattice site in a 12$\times$12 supercell of monolayers of 
(a) MoSe$_2$ and (b) MoTe$_2$ without allowing the host atoms to relax. Displacing the V atom along the dashed black lines in (b) weakens the V-Te$_1$ bonds while strengthening the V-Te$_2$ and V-Te$_3$ bonds. The brown circles in (b) represent the projection of the Te positions onto the metal plane. 
(c) Band structure without spin-polarization of the V-doped MoTe$_2$ monolayer for a 12$\times$12 supercell with JT distortion corresponding to one of the minima in (b).
The contributions from the vanadium $d_{xy}$, $d_{x^2-y^2}$ and $d_{z^2}$ orbitals are shown as half-filled green, half-filled blue and open red circles, respectively. Partial charge densities of the bands from (c) with $d_{xy}$ (d) and $d_{x^2-y^2}$ (e) character with an isovalue of 0.001 e \AA$^{-3}$. The red arrow indicates the displacement of the dopant atom (purple circle) when relaxed.  
}
\end{figure*}
% ====================

A substitutional impurity with partly filled, orbitally degenerate gap states satisfies the requirements of the Jahn-Teller (JT) theorem \cite{Jahn:prsa37, Inui:96} and should undergo a symmetry-lowering distortion at sufficiently low temperature \cite{Lannoo:81, Bourgoin:83}. The distortion will split the partly-filled orbitally-degenerate level  leading to a gain of single-particle energy that is linear in the amplitude $x$ of the distortion whereas the elastic restoring force will be quadratic in $x$ \cite{footnote1}. In equation form the JT energy $E_{\rm JT}$ can be written as  
\begin{equation}
\label{eq:JT}
  E_{\rm JT}(x)   = -F_{\rm JT} \, x+ \frac{1}{2}Kx^2            \\
\end{equation}
where $F_{\rm JT}$ is the force driving the JT distortion, $F_{\rm JT}\,x$ is the energy gained by the symmetry lowering, and the elastic restoring force $-Kx$ depends on the strength of the bond between the dopant atom and the neighbouring chalcogen atoms. In equilibrium, $x_0= F_{\rm JT}/K$ and the energy lowering is $\Delta E(x_0)=-F_{\rm JT}^2/2K=-x_0*F_{\rm JT}/2$. If the harmonic response of the bonds to the displacement is stiff (large elastic constant $K$) and the $e'$ orbitals are very delocalized (small $F_{\rm JT}$), the JT distortion may be very weak resulting in a negligibly small displacement $x_0$.  
The situations for single (a) and double (b) acceptors with one, respectively, two holes in the $e'$ level are sketched schematically in \Cref{fig:JT}; the $a'_1$ level is not shown. The displacement $x$ leads to a splitting $\delta$ of the partly filled $e'$ level that is linear in $x$.
An exchange splitting $\Delta$ of the orbitally degenerate level that can compete with the symmetry lowering is included \cite{footnote3}.
The distortion relevant for a substitutional impurity is the displacement of the dopant atom from the center of an MX$_6$ trigonal prism indicated in \Cref{fig:JT_structure} with a red arrow; this lowers the original local point group symmetry from $D_{3h}$ to C$_{2v}$. The dopant atom moves in the metal plane so mirror symmetry with respect to this plane is preserved. 

For MoS$_2$, the small value of $\Delta_{K\Gamma}$ (\Cref{TableI}) makes the $a'_1$ impurity levels bound to the $\Gamma$ VBM accommodate the hole. Explicit tests in reference \cite{Gao:prb19b} showed that symmetry-lowering distortions were energetically unfavourable for single-acceptor doped MoS$_2$ monolayers and only symmetric relaxation was found to occur. In view of the near degeneracy of the $a'_1$ and $e'$ levels and the finite dispersion of these states when modelled in a finite supercell, we could understand that an eventual energy gain in the single impurity limit might be masked by the supercell approximation. 
However, for MoSe$_2$ and MoTe$_2$ monolayers doped with single acceptors, the $a'_1$ defect level lies well below the $e'$ gap state, \Cref{TableI}. The latter contains a single hole and is therefore susceptible to a Jahn-Teller distortion \cite{Jahn:prsa37, Inui:96}. However, no symmetry-lowering distortion is found for MoSe$_2$:V (vanadium-doped MoSe$_2$) while it does reduce the total energy for MoTe$_2$:V. To better understand what is happening, we examine these two cases in more detail by constructing the potential energy surfaces shown in \Cref{fig:PES} for (a) MoSe$_2$:V and (b) MoTe$_2$:V monolayers by calculating the total energy of a dopant V atom as a function of its displacement in the $xy$ metal plane in a $12\times12$ supercell without allowing the host atoms to relax. 

For the MoSe$_2$:V monolayer the potential minimum corresponds to an undisplaced, symmetric V dopant. The situation is unchanged and the $D_{3h}$ symmetry  maintained when the neighbouring atoms are allowed to relax. The strong V-Se bond with significant hybridization of the V $d_{xy/x^2-y^2}$ and Se $p$ orbitals leads to a large elastic restoring force. The gain of electronic energy from the JT distortion fails to compensate the cost of elastic energy and no JT distortion is found for single-acceptor doped MoSe$_2$ monolayers in calculations with the largest supercell we considered ($12 \times 12$).  

By contrast, the potential energy surface for the MoTe$_2$:V monolayer has three equivalent in-plane minima with the V atom displaced away from a pair of nearest neighbour Te atoms above and below the Mo plane (Te$_1$ in \Cref{fig:PES}) weakening the corresponding V-Te bonds. The JT distortion corresponding to this displacement of the dopant atom from the high-symmetry site breaks the local $D_{3h}$ symmetry, lifts the degeneracy of the $e'$ levels with $d_{xy}$ and $d_{x^2-y^2}$ orbital character, \Cref{fig:PES}(c), and lowers the energy. From the orbital-resolved band structure for MoTe$_2$:V shown in \Cref{fig:PES}(c), we see that the impurity level with $d_{xy}$ orbital character is raised by the JT distortion while the impurity level with $d_{x^2-y^2}$ orbital character is lowered. 
The partial charge densities of the bands with V $d_{xy}$, \Cref{fig:PES}(d), and $d_{x^2-y^2}$, \Cref{fig:PES}(e), orbital character show that the JT displacement of V away from Te$_1$ reducing the V-Te$_2$ and V-Te$_3$ bond lengths increases the hybridization between the V $d_{xy}$ orbital and Te $p$ orbitals and raises the impurity levels with $d_{xy}$ orbital character. On the contrary, the band with $d_{x^2-y^2}$ orbital character is lowered by weakening the V-Te$_1$ antibonding interaction. As illustrated in \Cref{fig:JT}(a), the JT distortion does not quench the magnetic moment but the symmetry breaking of the local structure does affect the exchange interaction between pairs of dopant atoms. This will be discussed in \Cref{sec:disc}. 

For V-doped MoTe$_2$ monolayers, the V$-$Te bond is weaker than the V$-$Se bond \cite{Tariq:jms21} and our first-principles calculations yield a JT displacement of $x_0 =0.04$ \AA. The energy gained by the JT distortion is determined from the difference in energies of the JT distorted structure and of the same structure with the dopant returned to its initial high symmetry position. The net energy gain is found to be $\Delta E(x_0)=7$ meV. Using the simple model described by \eqref{eq:JT} according to which $\Delta E(x_0)=-F_{\rm JT}^2/2K=-x_0*F_{\rm JT}/2$, we estimate $F_{\rm JT} \sim 0.014/0.04=0.35\,$eV/\AA\ and $K \sim 0.35^2/0.014 = 8.75 \,$eV/\AA$^2$. 

In the MX$_2$ compounds, the bond strength ($\sim$ elastic constant $K$) decreases as X goes from S to Te \cite{Michel:prb17, Tariq:jms21}, making it more advantageous for the dopant to move from its original position. The lattice constant of MoTe$_2$ is larger than that of MoSe$_2$ that is in turn larger than that of MoS$_2$ so the impurity atom is in a larger cavity and presumably becoming less and less comfortable with this. The JT driving force described by $F_{\rm JT}$ is the single-particle-energy gained by breaking the degeneracy of the $e'$ levels. We can understand our failure to observe a JT distortion for MoS$_2$:V and MoSe$_2$:V while finding one for MoTe$_2$:V in terms of the impurity bandwidth becoming smaller in the same order making the degeneracy-lifting easier. For an unrelaxed, undisplaced substitutional V, the bandwidth of the $e'$ impurity levels is found to decrease from 19.7~meV$\rightarrow$12.5~meV$\rightarrow$9.8~meV for S$\rightarrow$Se$\rightarrow$Te.  $F_{\rm JT}$ increases and $K$ decreases so that the JT distortion as described by \eqref{eq:JT} becomes more favourable in the same order.

To confirm the relationship between the impurity bandwidth and the JT distortion, we add an on-site Coulomb repulsion, $U-J=5\,$eV \cite{Dudarev:prb98} to the V-3$d$ orbitals in the V-doped MoSe$_2$ monolayer. This pushes electrons from the V onto the neighbouring atoms,  localizing the hole on the V. The effective Bohr radius $a^*_0$ is reduced from 6.9 \AA\ to 5.3 \AA, resulting in a reduced impurity bandwidth and an increased JT displacement of $x_0 =0.07$ \AA. 
Nb and Ta dopants behave just like V in that no JT distortion occurs for MoS$_2$ and MoSe$_2$ systems. The impurity bandwidths of the $e'$ states obtained for Nb and Ta in MoTe$_2$ using a 12$\times$12 supercell are larger than that found for V leading to a weaker JT distortion with a small displacement of only about 0.01 \AA. 

The modelling of single impurities with supercells is ultimately limited by the maximum supercell size that can be handled using the {\sc vasp} (or any other) plane-wave code. In the single impurity limit (infinite supercell size) the impurity levels become dispersionless and the localization of the impurity wavefunction increases the JT driving force facilitating the JT distortion; in the single impurity limit the JT theorem must hold at $T=0\,$K. 

Replacing S in MoS$_2$ with Se and Te leads to a substantial lowering of the $\Gamma$-point VBM with respect to the $K$-point VBM and a corresponding lowering of the defect state bound to the $\Gamma$ VBM. The same is found to hold true for the Cr and W diselenides and ditellurides as seen in the increased value of $\Delta_{K\Gamma}=\varepsilon(K)-\varepsilon(\Gamma)$ in \Cref{TableI} and in the rapid descent of the $a'_1$ states with respect to the $e'$ states in \Cref{fig:impuritylevel}. Just as we found for single acceptors in MoS$_2$, only a symmetric relaxation is found for single-acceptors in CrS$_2$ and WS$_2$. For CrSe$_2$, CrTe$_2$ and WSe$_2$ monolayers, no JT distortions are found. In some systems which should be JT unstable, the large elastic constant and delocalized impurity wavefunctions prevent the JT distortion occurring when these systems are modelled using 12$\times$12 supercells. 

%-------10--------20--------30--------40--------50--------60--------70--------80
\subsubsection{Jahn-Teller distortions: double acceptors}
\label{sssec:JTda}

In \Cref{sssec:BE} we noted that for double acceptors (rhs of \Cref{fig:impuritylevel}), the main trends for the $a'_1$ and $e'$ levels in MX$_2$ paralleled those found for single acceptors, the main difference being much larger impurity binding energies. For MS$_2$ monolayers, the $a'_1$ levels were placed well above the $e'$ levels and were occupied with the two holes introduced by Ti, Zr and Hf doping. As a result there is no JT distortion and the relaxation is similar to that found for single-acceptors. 
For MSe$_2$ and MTe$_2$ monolayers, the $a'_1$ levels follow the $\Gamma$-point VBM down in energy losing their holes to the $e'$ levels that, being half filled, are susceptible to Jahn-Teller distortion. 

As suggested in \Cref{fig:JT}(b), there is competition between JT distortion and spin polarization. If only symmetric relaxation is allowed with spin polarization, then the double acceptors polarize fully (rhs). However, if symmetry-lowering is allowed, then the JT distortion quenches the magnetic moment (lhs). This indicates that two local minima exist that are close in energy: a JT state without spin polarization (lhs) and a spin-polarized state without JT distortion (rhs).

Which solution is found depends on how the DFT calculations are performed. If we start with a symmetrical configuration with both holes in a fourfold orbitally- and spin-degenerate state and first allow it to spin-polarize, then the fourfold degenerate $e'$ level splits into twofold up-spin and down-spin levels. 
The uppermost $e'$ level is filled with the two holes and is no longer JT active; the dopant atom does not move and the local symmetry is not broken. 
If we start with an unpolarized state with a half filled spin-degenerate $e'$ level, the impurity atom moves off center and the uppermost $d_{xy}$ level is filled with an up-spin and a down-spin hole and cannot spin-polarize. 
Apart from the position of the dopant atom itself, the relaxed local atomic geometries with spin polarization (SP) and without (JT state) are almost identical. 
Assuming the same symmetric relaxation around the dopant and correspondingly, the same energy gain, then we can model the total energy as
\begin{equation}
\label{eq:JT1}
   E_{\rm SP}    = E_0 - E_{\rm sym} - E_\Delta           \\ 
\end{equation}
\begin{equation}
\label{eq:JT2}
    E_{\rm JT}(x_0)   = E_0 - E_{\rm sym} -F_{\rm JT} \, x_0 + \frac{1}{2}Kx_0^2 ,     \\
\end{equation}
respectively, in which $E_{\rm sym}$ and $E_\Delta$ represent the energy gain from the symmetric relaxation and exchange splitting, respectively. $E_0$ is the total energy without relaxation and spin polarization while $x_0$ represents the equilibrium displacement of the dopant atom.

% ========== Table 4 ==========%
\begin{table} [b]
\caption{Calculated net energy gain $\Delta E$ in meV of the JT state with respect to the spin-polarized state for double-acceptor doped MX$_2$ (M=Cr,Mo,W;X=Se,Te) monolayers. }
\begin{ruledtabular}
\begin{tabular}{llrrrrr}
& M   &  Cr & Mo &  W &  Cr & Mo  \\                                                  
 \cline{3-5} \cline {6-7} 
M$'$ & X   & \multicolumn{3}{c}{Se}
                    & \multicolumn{2}{c}{Te} \\
\hline
Ti  & & 54 & -45 & -10 & 42 & -90 \\
Zr  & & 70 & -11 &  -6 & 56 & -82 \\
Hf  & & 76 & -12 &  -5 & 59 & -80  
\end{tabular}
\end{ruledtabular}
\label{TableV}
\end{table}
% ========== Table 2 ==========% 

For single acceptors in MSe$_2$ and MTe$_2$ monolayers, the JT distortion does not compete with spin polarization as sketched in \Cref{fig:JT}. For double acceptors with two holes, only  singlet states are compatible with lifting of the orbital degeneracy of the $e'$ levels. The net energy gain of the JT state with respect to the spin-polarized state is 
\begin{equation}
\label{eq:JT3}
  \Delta E = E_{\rm JT}(x_0) - E_{\rm SP} = -F_{\rm JT} \, x_0 + \frac{1}{2}Kx_0^2 + E_\Delta.           \\
\end{equation}
The impurity states delocalize on going from Ti to Hf leading to a weaker driving force $F_{\rm JT}$ and so a smaller stabilization energy $\Delta E$ as listed in \Cref{TableV}. 
In double-acceptor doped MoSe$_2$, MoTe$_2$, and WSe$_2$ monolayers, the JT distorted state is more favorable than the spin-polarized solution with negative values of $\Delta E$. As discussed above, a MoTe$_2$ monolayer has a relatively small $K$ and large $F_{\rm JT}$ compared to a MoSe$_2$ monolayer. As a result, ($\Delta E$ and $x_0$) are quite large in a double-acceptor doped MoTe$_2$ monolayer: Ti (-90 meV, 0.11 \AA) and Hf (-80 meV, 0.11 \AA). A WSe$_2$ monolayer has a large elastic constant $K \sim$147 N/m \cite{Michel:prb17} (9.2 eV/\AA$^2$), resulting in a small value of $\Delta E$: -10 meV for Ti and -5 meV for Hf. It should be noted that the direction of the JT displacement for double acceptors is opposite to that of single acceptors.

The structural relaxation in doped CrX$_2$ and WX$_2$ monolayers is similar to that in MoX$_2$. In double-acceptor doped CrSe$_2$ and CrTe$_2$, the  $e'$ states are quite shallow (see rhs of \Cref{fig:impuritylevel}) and correspondingly delocalized which leads to a weak JT driving force $F_{\rm JT}$. The $a'_1$ levels that are close to the $e'$ levels broaden into impurity bands in the supercell approximation, see e.g. \Cref{fig:PES}(c), and become partially populated with holes. This increases $E_\Delta$ because of the larger exchange splitting of the $a'_1$ levels. When this happens, $\Delta E$ becomes positive indicating that spin-polarized states are energetically more favourable than JT states. 

%-------10--------20--------30--------40--------50--------60--------70--------80
\subsubsection{Single ion anisotropy}
\label{sssec:SIA}

For the isotropic 2D Heisenberg model with a gapless spin-wave spectrum, thermal fluctuations at any finite temperature will destroy long-range magnetic ordering whereas the magnetic anisotropy implicit in the 2D Ising model makes the system undergo a phase transition at a finite temperature to a magnetically ordered state \cite{Mermin:prl66, Hohenberg:pr67}; magnetic anisotropy is essential for long-range ferromagnetic ordering in two dimensions.
For the recently reported ferromagnetic CrI$_3$ \cite{Huang:nat17} and Fe$_2$GeTe$_3$ \cite{Fei:natm18, Deng:nat18} monolayers, single ion anisotropy (SIA) plays an important role in stabilizing the long-range order \cite{Xu:npjcm18, Fei:natm18}. We can expect the same to hold for MX$_2$ monolayers doped to make them magnetic, making it important to determine their magnetic anisotropy. We will study this in the single impurity limit. 

% ========== Figure 7 ==========
\begin{figure}[t]
\includegraphics[scale = 0.26]{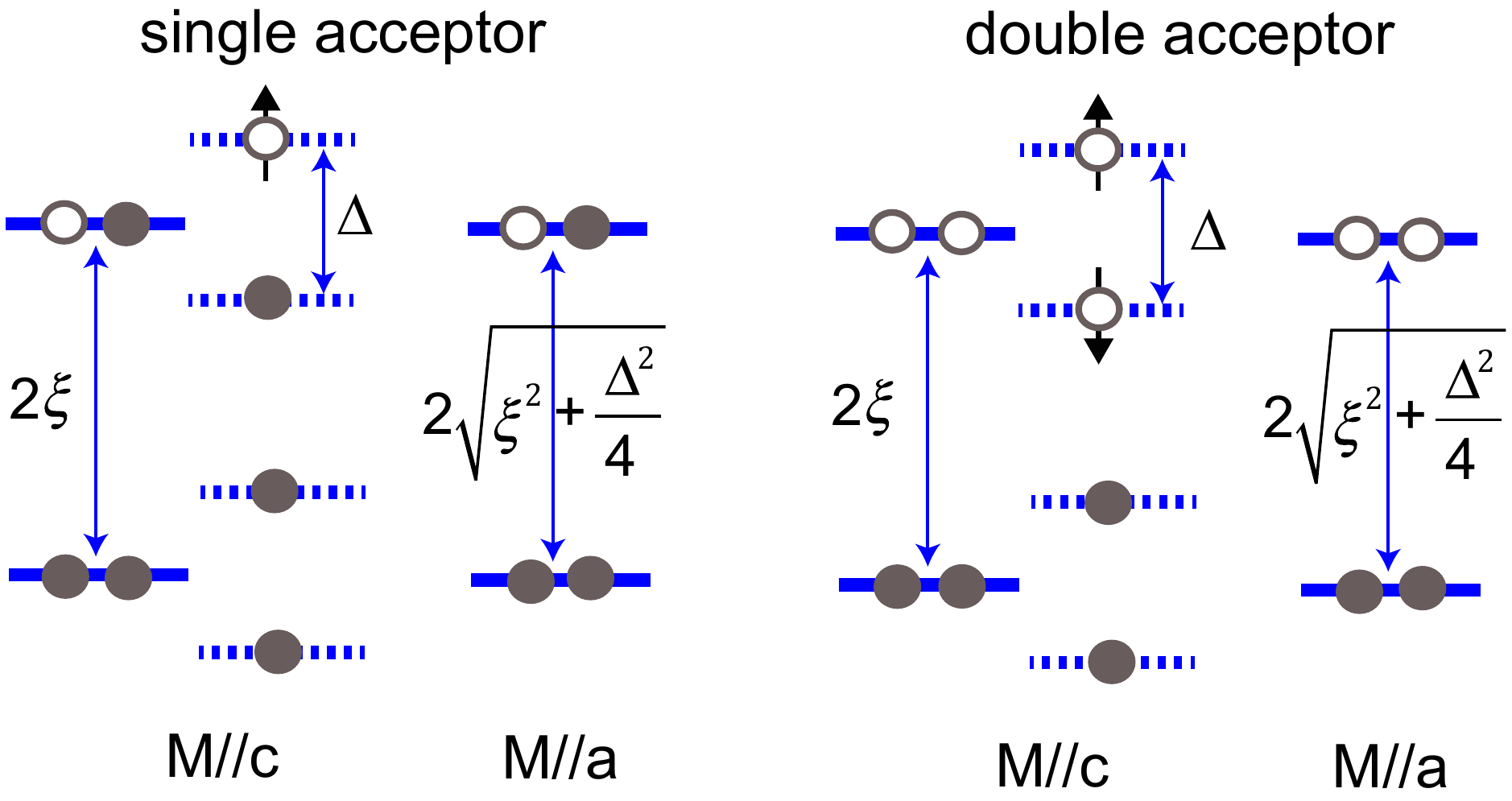}
\caption{\label{fig:MAE} Schematic diagram of the energy gain with spin orientation in-plane (${\bf M}\! \parallel \! a$) and out-of-plane (${\bf M}\! \parallel \! c$) for single-and double-acceptors. The solid and dashed blue lines represent spin degenerate and spin resolved levels, respectively. Solid and open circles denote electrons and holes, respectively.} 
\end{figure}
% ====================

To be able to interpret the SIA calculated from first-principles, we consider the model Hamiltonian for the doped system including spin orbit coupling (SOC) and spin polarization 
\begin{equation}
\label{MAE}
  H   = H_0 + \Delta {\bf m}.{\bf s}  + \xi {\bf l} \cdot {\bf s}
\end{equation}
in which $H_0$ is the spin-independent part of the Hamiltonian,
$\Delta {\bf m}$ is the exchange field that leads to an exchange splitting $\Delta$,
$\bf m$ is a unit vector in the direction of the magnetization, $\bf m \equiv M/|M|$, and $\xi$ is the conventional spin-orbit coupling parameter. In the subspace of the $l=2, m_l=\pm 2$ orbitals
\begin{equation}
 \xi {\bf l} \cdot {\bf s}
 = \frac{\xi}{2}  \begin{pmatrix} l_z & l_- \\   l_+ & -l_z  \end{pmatrix}
 =                \begin{pmatrix} \xi & 0   \\   0   & -\xi  \end{pmatrix}
\end{equation}
where we use Hartree atomic units with $\hbar=1$.
For ${\bf M} \! \parallel \! c$, 
\begin{equation}
\Delta {\bf m}.{\bf s}
 =                \begin{pmatrix} \frac{\Delta}{2}   & 0   \\   0   & -\frac{\Delta}{2}   \end{pmatrix}.
\end{equation}
and the SOC Hamiltonian can be written as
\begin{equation}
H   = H_0 + \begin{pmatrix}
  \xi + \frac{\Delta}{2} &      0                  &     0                 &  0 \\
      0                  & -\xi - \frac{\Delta}{2} &     0                 &  0 \\
      0                  &      0                  & -\xi+\frac{\Delta}{2} &  0 \\
      0                  &      0                  &     0                 & \xi-\frac{\Delta}{2} \\
                        \end{pmatrix}
\end{equation}
Rotating the spin quantization direction from $c$ to $a$ we get for ${\bf M} \! \parallel \! a$,
\begin{equation}
H   = H_0 + \begin{pmatrix}
      \xi        & \frac{\Delta}{2} &   0               &  0                \\
\frac{\Delta}{2} &      -\xi        &   0               &  0                \\
      0          &      0           &       -\xi        &  \frac{\Delta}{2} \\
      0          &      0           &  \frac{\Delta}{2} &        \xi        \\
                        \end{pmatrix}.
\end{equation}
Diagonalizing $H$ results in the energy level scheme sketched in \Cref{fig:MAE}. Occupying these levels as appropriate for single and double acceptors results in the total energies $E_c$ and $E_a$ for ${\bf M} \! \parallel \! c$ and  ${\bf M} \! \parallel \! a$, respectively, from which the single ion anisotropy energy $E_{\rm MAE}=E_a-E_c$ can  be estimated to be
\begin{equation}
\label{eqn:SIA}
E_{\rm MAE} = \begin{cases}
\xi + \frac{\Delta}{2} - \sqrt{\xi^2+\frac{\Delta^2}{4}}  & \mbox{for V, Nb, Ta} \\  
                                      &                      \\  
      2\xi -  \sqrt{4\xi^2+\Delta^2}  & \mbox{for Ti, Zr, Hf} 
\end{cases}
\end{equation}
where positive values favour an out-of-plane spin orientation and negative values in-plane. Equation \eqref{eqn:SIA} shows how the SIA depends on both the SOC and exchange interaction. It is clear that single acceptors yield positive values; double acceptors yield negative values (if the magnetic moments are not quenched by a JT distortion on iterating to self-consistency). 

We calculated the SIA numerically from first principles using the force theorem \cite{Mackintosh:80, Heine:80} which expresses the difference between two total energies as the difference in the sum of Kohn-Sham eigenvalues \cite{Kohn:pr65} for  ${\bf M} \! \parallel \! a$ and  ${\bf M} \! \parallel \! c$  \cite{Daalderop:prb90a, *Daalderop:prl92, *Daalderop:prb94}. \Cref{TableVI} compares the SIA from \eqref{eqn:SIA} with the first-principles results. We note that the spin-orbit splitting of the $e'$ levels for the V-doped MoS$_2$ monolayer is similar (130 meV) to that of the $K/K'$ VBM of a pristine MoS$_2$ monolayer, while the $a'_1$ level has no spin-orbit splitting just as the $\Gamma$ valley of the MoS$_2$ monolayer has none. The exchange splitting is 15.2 meV and 96 meV for the $e'$ and $a'_1$ levels, respectively. In \Cref{TableVI}, we see an estimate of 7.2 meV using \eqref{eqn:SIA}, larger than what we calculate from first principles, 4.5 meV. 
We note that because of the near degeneracy of the $a'_1$ and $e'$ levels and the impurity band dispersion in a supercell approximation, the $a'_1$ level may accommodate some of the hole. This does not contribute to the SIA, leading to the first-principles SIA being smaller than that estimated using \eqref{eqn:SIA} that does not take this factor into account. In addition, because the $e'$ level has become a band with a finite dispersion, the energy gain from the exchange splitting $\Delta$ will be reduced which will also reduce the SIA. The values from \eqref{eqn:SIA} are a simple estimate of the SIA in the single impurity limit with all holes in $e'$ states. For all of the systems in \Cref{TableVI} there is reasonable qualitative agreement between the DFT supercell calculations and the SIA estimate of equation \eqref{eqn:SIA}.  
 
% ========== Table V ==========%
\begin{table} [tbp]
\caption{Calculated spin orbit splitting $2\xi$, exchange splitting $\Delta$, single ion anisotropy (SIA) in meV, and effective Bohr radius $a^*_0$ in \AA\ for the degenerate $e'$ levels in monolayers of MX$_2$ ($12 \times 12$ supercell) doped with V and Ti substituting a single Mo. The SIA estimated from \eqref{eqn:SIA} is also shown for comparison. The atomic structures are fully relaxed with spin polarization.}
\begin{ruledtabular}
\begin{tabular}{lcccccccccc}
M &                    & Cr  &   Mo  &  W   &  Cr  &  Mo  &  W  & Cr &   Mo \\                                                  
\cline{3-5} \cline {6-8} \cline {9-10}
X &                    &  \multicolumn{3}{c}{S}
                            & \multicolumn{3}{c}{Se}
                                                    & \multicolumn{2}{c}{Te} \\
\hline
V  & $2\xi$               & 66   & 130   & 187   & 78   & 107   & 181   &  91   & 110   \\
   & $\Delta$             & 21.5 &  15.2 &   9.3 & 24.2 &  26.7 &  12.0 &  24.1 &  44.0 \\
   & SIA: \eqref{eqn:SIA} &  9.0 &   7.2 &   4.5 & 10.3 &  11.0 &   5.8 &  10.5 &  17.8 \\
   & SIA:DFT              &  4.7 &   4.5 &   3.2 &  7.2 &  10.0 &   4.7 &   7.7 &  11.2 \\
   & $a^*_0$              &  8.3 &   8.0 &   8.8 &  8.9 &  6.9 &   9.4 &   9.3 &  6.4  \\
 \hline
Ti & $2\xi$               &  0.0 &   0.0 &   0.0 & 78   &  87   & 148   &  83   & 105   \\
   & $\Delta$             &  0.0 &   0.0 &   0.0 & 13.3 &  84.8 &  64.4 &  71.4 & 90.2  \\
   & SIA: \eqref{eqn:SIA} &  0.0 &   0.0 &   0.0 & -1.1 & -34.4 & -13.4 & -26.4 & -31.1 \\
   & SIA:DFT              &  0.0 &   0.0 &   0.0 & -0.3 & -24.1 &  -5.2 & -23.4 & -25.5 \\
\end{tabular}
\end{ruledtabular}
\label{TableVI}
\end{table}
% ========== Table 1 ==========%

In the limit $\Delta \ll \xi, E_{\rm SIA} \sim \frac{\Delta}{2}(1-\frac{\Delta}{4\xi})$ for single acceptors and the SIA depends mainly on the exchange splitting $\Delta$. In MoSe$_2$, the SIA can be as high as 10 meV/dopant, larger than the 4.5 meV/dopant in MoS$_2$. V in MoTe$_2$ has a very large exchange splitting but because of the JT distortion, the SIA is not as high as the value estimated from \eqref{eqn:SIA}. Pristine WS$_2$ and  WSe$_2$ monolayers have a large spin-orbit splitting of the $K/K'$ valley ($\sim$ 0.4-0.6 eV), but the splitting of the impurity levels in these systems is only half of these values. The more delocalized the holes become, the smaller is $\Delta$ and consequently the SIA. In CrX$_2$ monolayers, the impurity states have a large exchange splitting $\Delta$ but the small SOC leads to SIA values intermediate between those found for MoS$_2$ and WS$_2$.

%%%%%%%%10%%%%%%%%20%%%%%%%%30%%%%%%%%40--------50--------60--------70--------80
\subsection{Interaction with intrinsic defects}
\label{ssec:ID}

Intrinsic defects like vacancies and antisite defects are ubiquitous in van der Waals materials and have been investigated in many experimental \cite{Hong:natc15, Wang:csr18, Rasool:am15} and computational \cite{Komsa:prb15, Li:prb16, Haldar:prb15} studies. We denote vacancies on M and X sites as $\O_{\rm M}$ and $\O_{\rm X}$, respectively. Analogously, antisite defects are denoted A$_{\rm B}$ for an A atom on a B site; thus in the present case we will be concerned with M$_{\rm X}$. The formation energies of the various defects listed in \Cref{TableVII} are defined as
\begin{equation}
\label{eq:form-va}
E_{\rm form}[{\rm def}] = E_{\rm tot}[{\rm MX_2\!:def}] 
- \Big(E_{\rm tot}[{\rm MX_2}] - \sum_{\alpha} n_{\alpha} \mu_{\alpha}\Big) \\
\end{equation}
where $E_{\rm tot}[{\rm MX_2\!:def}]$ and $E_{\rm tot}[{\rm MX_2}]$ denote total energies of supercells containing a defect and of the perfect material, respectively; $n_\alpha$ is the number of atoms added or removed and $\mu_\alpha$ ($=\mu_{\rm M}$ or $\mu_{\rm X}$) represents the corresponding M or X chemical potential in their bulk phases (bcc or hexagonal phase for M atoms; monoclinic or triclinic phase for X atoms). The formation energy of a chalcogen X atom vacancy, $E[\O_{\rm X}]$, is about half that of a transition metal vacancy, $E[\O_{\rm M}]$, or half that of an antisite defect with a metal atom on an X site, $E[\rm M_X]$, indicating that chalcogen vacancies are more likely to be formed in thermodynamic equilibrium than the other two defects. The formation energies $\O_{\rm M}$ and M$_{\rm X}$ are fairly close in value and decrease from MS$_2$ to MTe$_2$. Compared to the formation energy of acceptors (see \Cref{TableII}), much more energy is required to form intrinsic defects. This suggests that in order to suppress the formation of intrinsic defects, acceptors should be incorporated into MX$_2$ monolayers at low temperatures. 
%In the following, we will focus on the polarization states introduced or affected by these intrinsic defects in $p$-doped systems. {\color{red}Ferromagnetism is an ordering phenomenon ... it's perhape better not to talk too much about the magnetism of atoms, defects etc.}

% ========== Table VII ==========%
\begin{table} 
\caption{Calculated formation energy in eV of a chalcogen vacancy $\O_{\rm X}$, a transition metal vacancy $\O_{\rm M}$, and an M$_{\rm X}$ antisite (one transitional metal atom occupying a chalcogen atom site) in a MX$_2$ monolayer including relaxation. }
\begin{ruledtabular}
\begin{tabular}{lcccccccc}
M   &   Cr  &   Mo  &   W   &   Cr  &   Mo  &   W   &   Cr  &   Mo  \\                                                  
\cline{2-4} \cline {5-7} \cline {8-9}
X   &  \multicolumn{3}{c}{S}
                            & \multicolumn{3}{c}{Se}
                                                    & \multicolumn{2}{c}{Te} \\
\hline
$\O_{\rm X}$ & 2.47 & 3.15 & 3.01 & 2.46 & 2.77 & 2.72 & 2.26 & 2.89  \\
$\O_{\rm M}$ & 6.31 & 7.87 & 7.27 & 4.80 & 6.04 & 5.85 & 2.44 & 4.60  \\
M$_{\rm X}$ & 4.72 & 6.71 & 7.12 & 4.44 & 5.51 & 6.48 & 3.31 & 4.61  \\
\end{tabular}
\end{ruledtabular}
\label{TableVII}
\end{table}
% ========== Table VII ==========% 

% ========== Figure 8 ==========
\begin{figure*}[t]
\includegraphics[width=17.6cm]{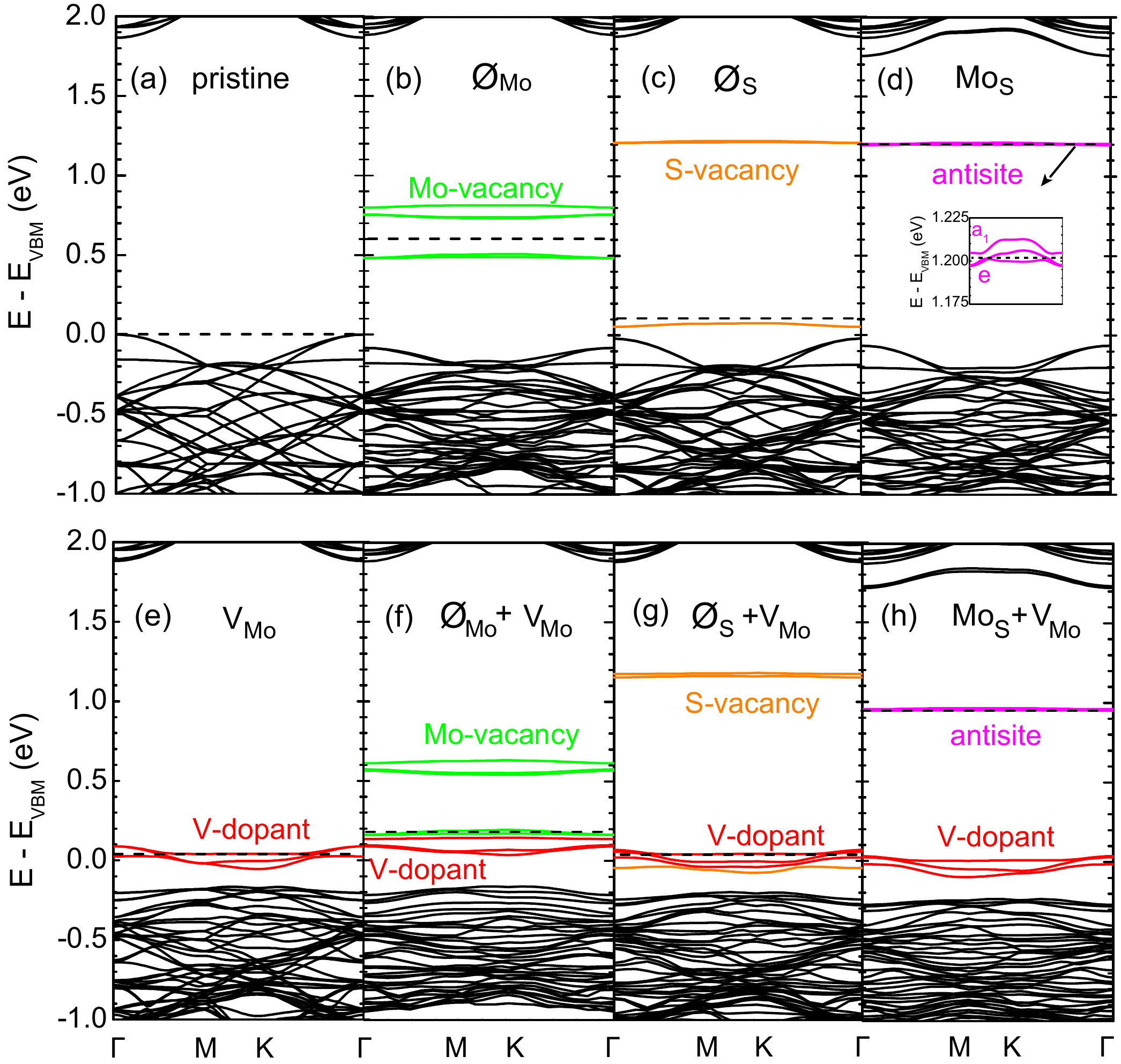}
\caption{\label{fig:defect_band} Band structures for 
	(a) a defect-free MoS$_2$ monolayer, 
	(b) a Mo vacancy, $\O_{\rm Mo}$, 
	(c) a sulfur vacancy, $\O_{\rm Mo}$,
	(d) a Mo$_{\rm S}$ antisite defect consisting of a Mo atom on a S site,
	(e) a vanadium atom on a Mo site, V$_{\rm Mo}$,
	(f) a Mo vacancy next to a substitutional vanadium atom, $\O_{\rm Mo}$+V$_{\rm Mo}$
	(g) a sulfur vacancy next to a substitutional vanadium atom, $\O_{\rm S}$+V$_{\rm Mo}$ 
	(h) a Mo$_{\rm S}$ antisite defect next to a substitutional vanadium atom, Mo$_{\rm S}$+V$_{\rm Mo}$
	all modelled for MoS$_2$ with a 6$\times$6 supercell. The zero of energy is the top of the valence band which is not always visible in a finite supercell; its determination is discussed in \cite{Gao:prb19b}. The Fermi level is indicated by a dashed black line. The inset in (d) is a blow-up of the band structure near the Fermi level.}
\end{figure*}
% ====================

Just as substituting a Mo atom with an acceptor (single or double) leads to the formation of states in the fundamental band gap, \Cref{fig:impuritylevel} and \Cref{fig:defect_band}(e), defect states associated with a variety of intrinsic defects appear in the band gap and can interact with those introduced by $p$ doping. This is illustrated in \Cref{fig:defect_band} for an MoS$_2$ monolayer where the non spin-polarized band structures for a number of intrinsic defects and complexes of  intrinsic defects and adjacent substitutional V dopant atoms, V$_{\rm Mo}$, are plotted. For simplicity, the structures are not relaxed and the effect of relaxation is discussed in the text.

A Mo vacancy, $\O_{\rm Mo}$, with a formation energy of 7.87 eV introduces five mid-gap states, one singly degenerate state and two doubly degenerate states under local $D_{3h}$ symmetry as shown in \Cref{fig:defect_band}(b). Six holes are hereby introduced into the system that occupy the three defect states above the Fermi level. Structural relaxation preserves the $D_{3h}$ symmetry as expected from the absence of partially occupied orbitally degenerate states. However, the relative position of the deep defect states depends on the particular system under consideration. For example, for a MoTe$_2$ monolayer with a Mo vacancy, the singly degenerate state lies far below the doubly degenerate states so that the Fermi level lies in the doubly degenerate states leading to a JT distortion.  

A sulfur vacancy, $\O_{\rm S}$, with a formation energy of 3.15 eV, introduces three gap states: an orbitally doubly degenerate, unoccupied midgap state and a fully occupied  singly degenerate state close to the VBM, \Cref{fig:defect_band}(c). Usually a sulfur vacancy is considered to be an electron donor but for that to be true, the occupied state would have to be close to the CBM which is not the case. There are no partially filled defect states so that structural relaxation preserves the $C_{3v}$ site symmetry. The Mo atoms surrounding the S vacancy move towards it which raises the occupied $\O_{\rm S}$ defect state slightly in energy. This is true for all MX$_2$ monolayers and chalcogen vacancies do not introduce spin-polarization or induce JT distortions.

A Mo$_{\rm S}$ antisite consists of a sulfur vacancy occupied by an extra Mo atom. The S vacancy removes two holes while the Mo atom contributes six electrons, \Cref{fig:defect_band}(d). Four of these electrons are accommodated in the sulfur vacancy mid-gap states that are pulled down into the valence band by the attractive Mo potential leaving two electrons in a deep doubly degenerate $e$ state under the local $C_{3v}$ symmetry that is quasi degenerate with an $a_1$ singlet and giving rise to a magnetic moment of $2 \mu_{\rm B}$. The inset of \Cref{fig:defect_band}(d) is a blow-up of the band structure near the Fermi level. For the partially occupied degenerate $e$ levels, JT distortion could lift the degeneracy but is not strong enough to do so and spin polarization prevails. The exchange-splitting of the doubly degenerate antisite states is 0.33 eV. For an MoTe$_2$ monolayer, relaxation lowers the $a_1$ level well below the $e$ levels. The $a_1$ level accommodates both electrons so that an Mo$_{\rm Te}$ antisite is nonmagnetic.

As discussed in detail in Ref.~[\onlinecite{Gao:prb19a, Gao:prb19b}], a vanadium atom on a Mo site, V$_{\rm Mo}$, induces three shallow impurity-like levels above the VBM which are shown unrelaxed in \Cref{fig:defect_band}(e). With relaxation, we see these levels with binding energies of about 60 meV in \Cref{fig:impuritylevel}. The intrinsic defect states shown in \Cref{fig:defect_band}(b-d) have very deep gap states ($>$ 0.5 eV), indicating that the defect potential is effectively more repulsive for holes than the substitutional V dopant. If we introduce a Mo vacancy next to a V dopant ($\O_{\rm Mo}$+V$_{\rm Mo}$), the V$_{\rm Mo}$ hole will transfer to the S atoms surrounding the Mo vacancy. As shown in \Cref{fig:defect_band}(f), the hole goes into the lowest lying doubly degenerate $\O_{\rm Mo}$ defect state and polarizes fully; this might suppress the quenching of the magnetic moment we found for close V dopant pairs in Ref.~[\onlinecite{Gao:prb19a, Gao:prb19b}]. 

A sulfur vacancy, $\O_{\rm S}$, has one fully occupied defect state that is very close to the VBM, \Cref{fig:defect_band}(c). When $\O_{\rm S}$ and V$_{\rm Mo}$ defects are neighbours, the shallow quasidegenerate defect states hybridize leading to a spatial redistribution of the hole, see \Cref{fig:defect_band}(g). It should be noted that the defect states of $\O_{\rm S}$ mainly consist of Mo $d$ states on the three Mo atoms that surround the S vacancy. The hybridization between V and Mo $d$ orbitals enhances the magnetism as evidenced by an increased exchange splitting. When allowed to relax, the two Mo atoms and the V dopant adjacent to the S vacancy move towards the vacancy site. This inward shift enhances the hybridization of the V$_{\rm Mo}$ and $\O_{\rm S}$ defect states increasing the exchange splitting of the corresponding $a'_1$ levels in a $6 \times 6$ supercell from $72\,$meV to $93\,$meV.

For a vanadium atom on a Mo site next to a Mo atom on a S site, a Mo$_{\rm S}$+V$_{\rm Mo}$ complex defect, the hole introduced by V$_{\rm Mo}$ moves to the partially filled Mo$_{\rm S}$ mid-gap states, resulting in a net magnetic moment of $1 \mu_{\rm B}$ on the antisite Mo atom. The V dopant gets one extra electron from the antisite defect, which quenches its magnetic moment. When relaxation is included, a JT distortion appears for the partially filled degenerate defect states of the antisite but the complex remains spin polarized.

All of the intrinsic defects we have examined are quite localized. Even though the local spin polarization can be enhanced by hybridization between intrinsic defects and dopant states in the sense that the exchange splitting is increased, this is unlikely to extend the range of the exchange interaction between magnetic defects. It has been reported that the existence of S vacancies might reduce the barrier for dopant diffusion \cite{Yang:prl19}. This might allow dopants to migrate and more easily form pairs in the presence of S vacancies, which could lead to an enhanced ordering temperaute at low doping concentration.

%%%%%%%%10%%%%%%%%20%%%%%%%%30%%%%%%%%40--------50--------60--------70--------80
\subsection{Electron doping: Rhenium}
\label{ssec:ED}

% ========== Figure 9 ==========
\begin{figure}[t]
\includegraphics[width=8.6cm]{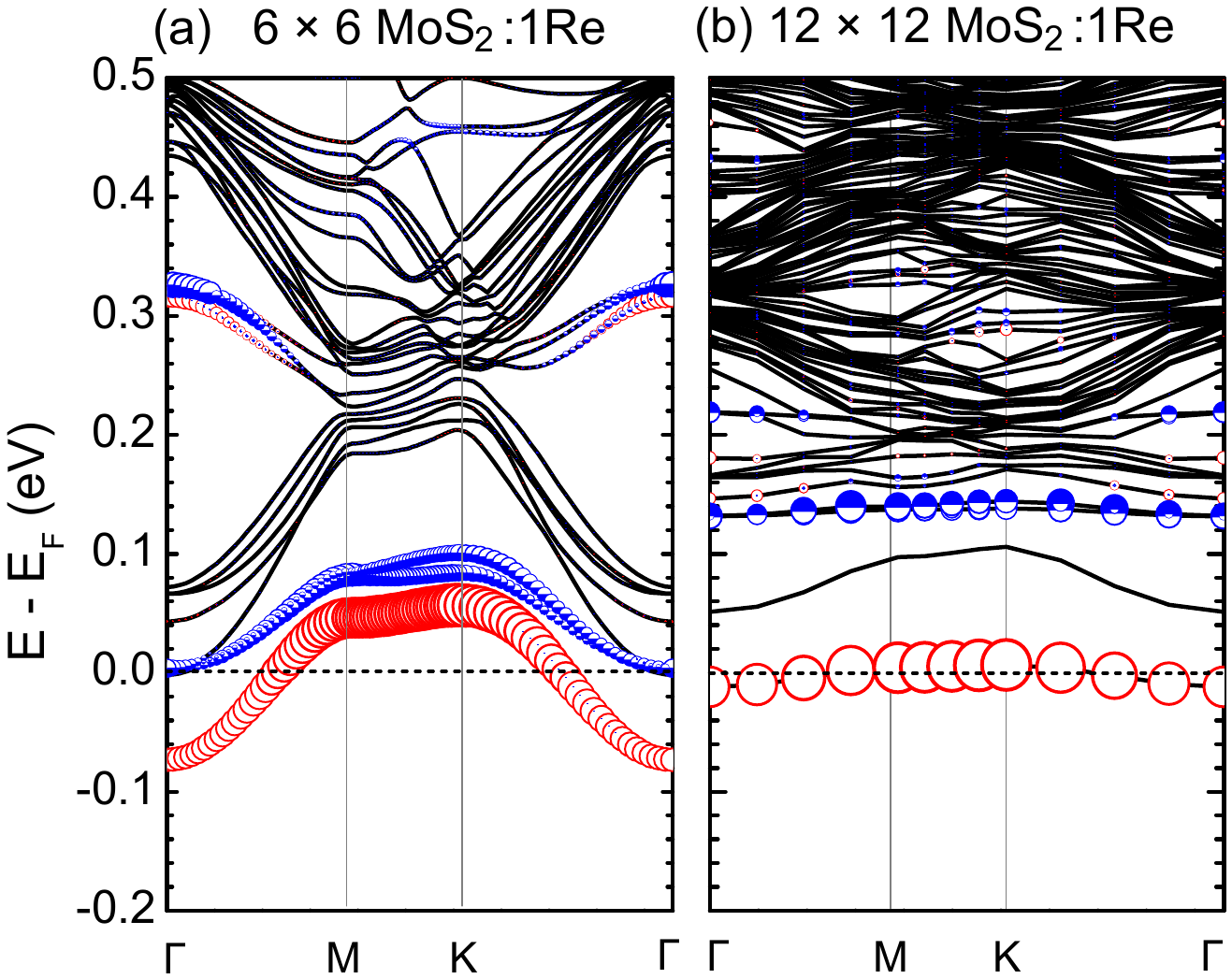}
\caption{\label{fig:Re}Orbital resolved band structure of an MoS$_2$ monolayer with one Mo atom replaced with a Re atom in a (a) 6$\times$6 and (b) 12$\times$12 supercell. The open red circles and half-filled blue circles represent the contributions from Re $d_{3z^2-r^2}$ and ${d_{x^2-y^2},d_{xy}}$, respectively. The Fermi level is indicated by the black dashed line.}
\end{figure}
% ====================

So far we have only considered the case of hole doping. Here we briefly consider the case of electron doping with the $5d$ element rhenium that gives rise to an attractive Coulomb potential in the single impurity limit. We expect the electron-doping case to be more complex because according to \Cref{fig:band} the conduction band minima (CBM) at $K/K'$ will give rise to two degenerate states while the six $Q$ point minima between $K$ and $\Gamma$ will give rise to an additional six degenerate states that will be coupled in degenerate perturbation theory by the Coulomb potential. In the effective mass approximation, the intervalley scattering between the $K$ and $K'$ valleys and the sixfold $Q$ valleys leaves a single $a'_1$ level with $d_{3z^2-r^2}$ character and a doubly degenerate $e'$ level with mixed $\{d_{x^2-y^2},d_{xy}\}$ character below the CBM as indicated in \Cref{fig:Re} with a binding energy of 92 meV. The lowest lying $a'_1$ state polarizes fully in the single impurity limit. In spite of its out-of-plane $d_{3z^2-r^2}$ character, the exchange splitting is weak (46 meV). This is manifested in the large dispersion of the $a'_1$ level seen in \Cref{fig:Re}. The small SOC at the CBM of MoX$_2$ monolayers leads to a negligible SIA. For WX$_2$ monolayers, it is relatively large compared to MoX$_2$ \cite{Jin:prl13, Wilson:sca17, Tanabe:apl16, Yeh:prb15} leading to a large SIA of $\sim$ 10 meV/dopant. However, as we found for pairs of substitutional V atoms in a MoS$_2$ monolayer \cite{Gao:prb19a, Gao:prb19b}, below a critical separation the magnetic moment is quenched by a strong $\pi$ bond being formed between Re pairs.

%%%%%%%%10%%%%%%%%20%%%%%%%%30%%%%%%%%40%%%%%%%%50%%%%%%%%60%%%%%%%%70%%%%%%%%80
\section{Results: Pair Interactions}
\label{sec:ResPI}

For semiconductors like the MX$_2$ transition metal dichalcogenides, the electronic structure of substitutional dopants from neighbouring columns of the periodic table can be qualitatively understood within the framework of effective mass theory. We have confirmed this for M substitutions using DFT-based supercell calculations once the supercell size is sufficiently large to approach the single impurity limit. For MoS$_2$ and WS$_2$ monolayers doped with single acceptors, the small energy difference $\Delta_{K\Gamma}$ between the $K/K'$ and $\Gamma$ valence band maxima means that the hole occupies the $a'_1$ level bound to the $\Gamma$ valley. For pairs of dopant atoms that are sufficiently close, the magnetic moment is quenched by the formation of singlet bonding states \cite{Gao:prb19b}. 
This quenching can be avoided by instead using MSe$_2$ and MTe$_2$ with M=Cr, Mo, W which have larger values of $\Delta_{K\Gamma}$, see \Cref{TableI}; even the slightly larger value of $\Delta_{K\Gamma}$ for CrS$_2$ is enough to suppress it. Nevertheless, in order to realize long range and strong exchange interactions at high temperatures, the impurity states should have large effective Bohr radii as well as large exchange splittings, requirements which are incompatible. The SIA of single acceptors also requires a large exchange splitting while double acceptors are excluded because of their in-plane anisotropy. In this section, we briefly consider the opportunities for realizing ferromagnetic ordering in the vanadium doped systems considered in the previous sections. 

Increasing the concentration of V doping increases the impurity bandwidth so the holes become more delocalized and the exchange splitting and MAE are reduced. This makes it more difficult for the Hund's-rule like alignment of spins to win the competition with the kinetic energy or for the MAE to win against thermal disorder. 
The effective Bohr radii $a^*_0$ of the $e'$ impurity states resulting from doping MX$_2$ with vanadium that are given in \Cref{TableVI} should be a measure of the range of the exchange interaction. The small values for MoSe$_2$ and MoTe$_2$ argue against their suitability as room temperature ferromagnets. 
If we use the limiting case of the single site exchange splitting $\Delta$ as a measure of the robustness of the magnetic moment to the adverse effects of temperature \cite{footnote3}, then \Cref{TableVI} shows that V-doped CrS$_2$, CrSe$_2$, CrTe$_2$, MoSe$_2$, and MoTe$_2$ have reasonably large values. Small values of $\Delta$ make the stabilization of parallel and antiparallel configurations of two spins difficult (and computationally expensive) in a SCF calculation and we do not consider WS$_2$ or WSe$_2$ any further; MoS$_2$ was examined in detail in \cite{Gao:prb19a, *Gao:prb19b}. 
(In terms of its Bohr radius, WSe$_2$:V might look very promising and indeed it has been reported to exhibit ferromagnetic ordering at room temperature \cite{Yun:advs20, Pham:am20} but not in terms of its intraatomic or interatomic exchange interactions that are weak.
We will discuss it separately in \Cref{sec:disc}.)  

% ========== Figure 10 ==========
\begin{figure}[t]
\includegraphics[width=8.0cm]{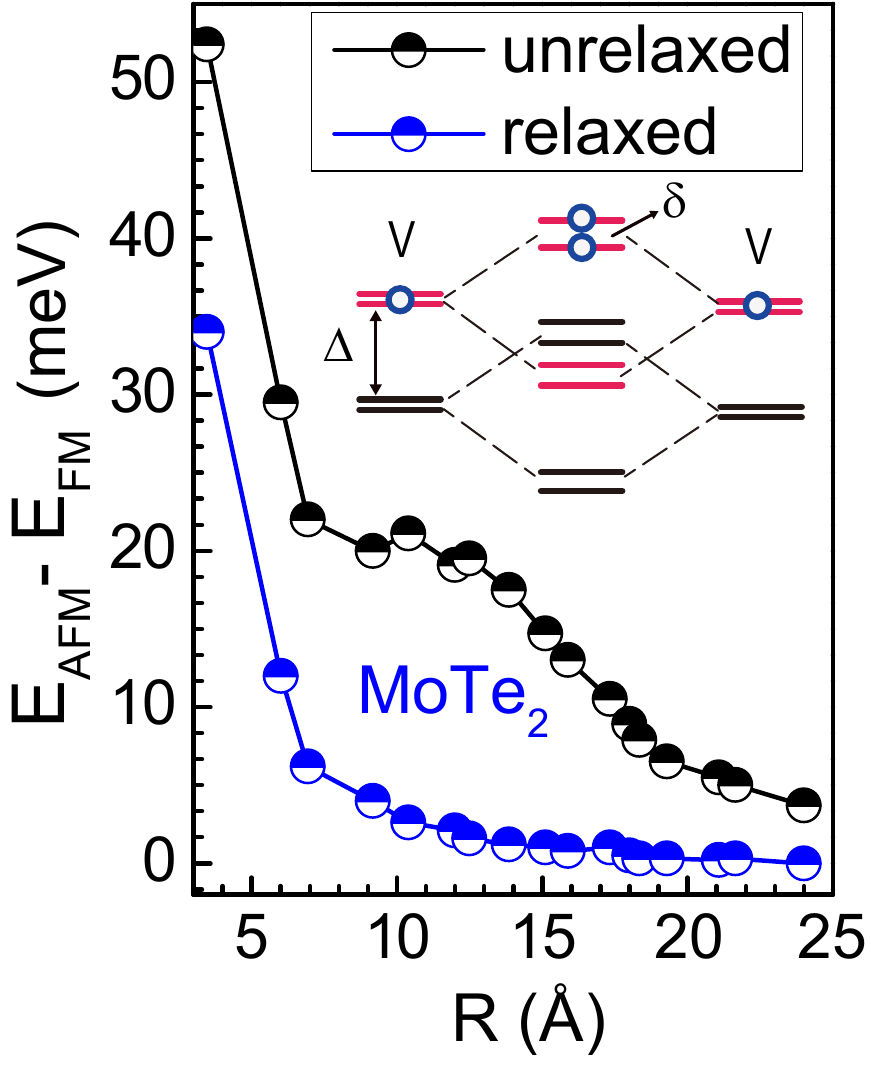}
\caption{\label{fig:MoTe-exchange}Difference between the total energies of antiferromagnetic and ferromagnetic states for pairs of substitutional V impurities in 12$\times$12 supercells plotted as a function of their separation for MoTe$_2$ monolayers with (half-filled blue circles) and without (half-filled black circles) relaxation. 
The solid lines are a guide to the eye. 
The inset shows the spin resolved bonding anti-bonding levels of ferromagnetically coupled V pairs in which red and black lines represent different spin channels. $\Delta$ represents the exchange splitting of the impurity states and $\delta$ the splitting of the degenerate $e'$ levels as the V impurities get closer and interact. 
Open circles denote the holes. 
}
\end{figure}
% ====================

Of the TMD monolayers considered in \Cref{TableVI}, V-doped MoTe$_2$ has the largest exchange splitting with $\Delta=$44 meV and we expect the single site magnetic moment to be robust against the influence of temperature and the interaction with other dopant impurities when the V concentration is increased. 
To gauge the latter, we consider the exchange interaction between pairs of V dopants separated by lattice vectors ${\bf R}$ by calculating the total energy difference between ferromagnetically and anti-ferromagnetically ordered states of V pairs in 12$\times$12 supercells, with and without relaxation as shown in \Cref{fig:MoTe-exchange}. By mapping this energy difference onto a Heisenberg model, Monte-Carlo calculations can be used to estimate ordering temperatures \cite{Gao:prb19a, *Gao:prb19b}. 

In the absence of relaxation, the hole of each individual V impurity occupies an orbitally doubly degenerate $e'$ state, \Cref{fig:impuritylevel}, that is exchange split by the amount $\Delta$ when spin-polarization is allowed \cite{footnote3}, as sketched in the inset to \Cref{fig:MoTe-exchange}.
Pairs of these V impurities exhibit a strong and long-range FM exchange coupling with points of inflection at separations of $\sim$7\AA\ and $\sim$14\AA\ between which the exchange energy flattens out, \Cref{fig:MoTe-exchange}. 
At large separations, the doubly degenerate $e'$ states form bonding-antibonding pairs with the holes gravitating to the highest-lying antibonding level (see the inset to \Cref{fig:MoTe-exchange}); because of the orbital degeneracy, the spins can form a triplet corresponding to ferromagnetic coupling.
Bringing the impurity pairs closer breaks the $C_{3v}$ rotational symmetry upon which the twofold degeneracy of the $e'$ levels is based and leads to a splitting $\delta$ of these levels. 
For separations between 14 \AA\ and 7 \AA, $\delta$ increases from approximately 3 to 7 meV. 
This leads to a reduction in the exchange interaction between V pairs which is evident in the formation of a plateau at separations between approximately 7 \AA\ and 14 \AA. 
At short separations, the bonding-antibonding interaction between  $d_{xy}$ and $d_{x^2-y^2}$ orbitals on V neighbours and the exchange splitting $\Delta$ of the impurity states are much larger than $\delta$, the symmetry breaking of the $e'$ level. Thus, as the dopants get closer ($<$ 7 \AA), the exchange interaction increases steeply.

When relaxation is included, the JT distortion lifts the degeneracy of the partly filled $e'$ levels as shown in \Cref{fig:PES}(c) suppressing the preference for  triplet states. As well as leading to the reduced MAE we found in \Cref{sssec:JTsa}, this greatly reduces the interatomic exchange interaction making it effectively much shorter range than expected on the basis of the Bohr radius for a single dopant impurity. One possibility to circumvent this would be to consider doping a Mo(Se$_x$Te$_{1-x})_2$ substitutional alloy with V; this however goes beyond the scope of the present manuscript. 

% ========== Figure 11 ==========
\begin{figure}[t]
\includegraphics[width=8.6cm]{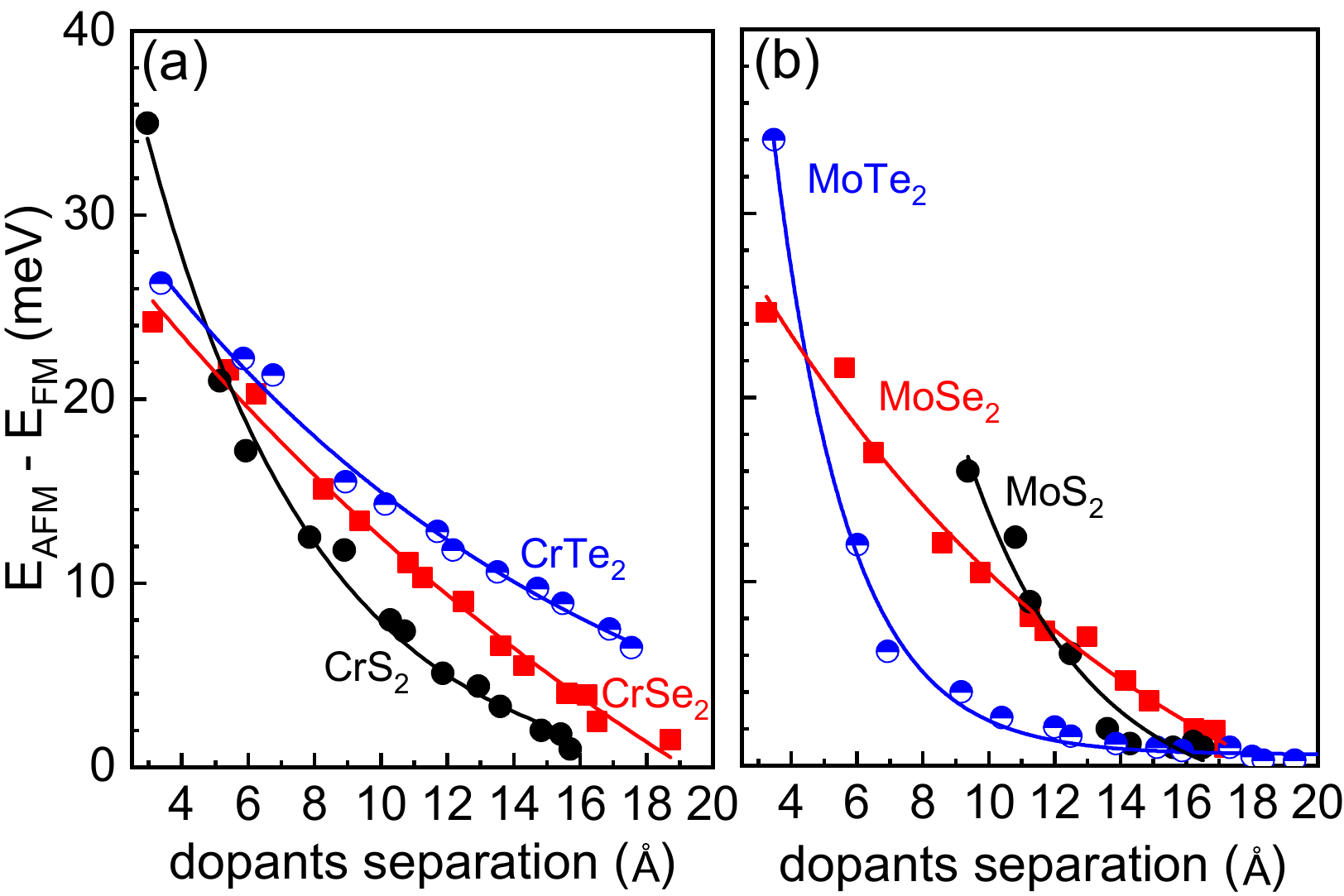}
\caption{\label{fig:exchange}Difference between the total energies of antiferromagnetic and ferromagnetic states for pairs of substitutional V impurities including relaxation in 12$\times$12 supercells plotted as a function of their separation for (a) CrX$_2$ and (b) MoX$_2$ (X= S, Se, Te) monolayers. The solid lines are a guide to the eye. 
}
\end{figure}
% ====================

The interatomic exchange interaction in the MoX$_2$ system, shown in \Cref{fig:exchange}(b), is quite complex because of (i) the magnetic quenching in MoS$_2$ discussed at length in reference~\cite{Gao:prb19a, *Gao:prb19b} and (ii) the JT effect in MoTe$_2$ discussed in the previous paragraph. While the JT distortion makes the exchange interaction of MoTe$_2$ very short-ranged, in MoS$_2$ the magnetic moment is quenched for V pairs below a critical separation ($\sim$ 9 \AA). In the intermediate Se case, our calculations found the JT distortion to be negligible for a single impurity and MoSe$_2$:V shows a relatively strong and long-range exchange interaction. 

For relaxed V pairs in CrX$_2$ monolayers, the ferromagnetic-antiferromagnetic energy difference is plotted as a function of the separation $R$ in \Cref{fig:exchange}(a) where the exchange interaction is seen to decay more slowly and gets stronger in the sequence CrS$_2$ $\rightarrow$ CrSe$_2 \rightarrow$ CrTe$_2$. 
Even though the Bohr radius of $e'$ levels in V-doped CrS$_2$ is 8.3 \AA, the exchange interaction drops faster with dopant separation than in MoSe$_2$ which has a smaller effective Bohr radius. This is because the $a'_1$ level in V-doped CrS$_2$ is so close to the $e'$ levels (see \Cref{fig:impuritylevel}) that are therefore only partly occupied, as we originally found for V-doped MoS$_2$ \cite{Gao:prb19b}. A key difference is that the $a'_1$ level in the V-doped CrS$_2$ monolayer is not high enough to quench the magnetic moment for even the closest dopant pair separation. It should be noted that in CrS$_2$, CrSe$_2$, CrTe$_2$ and MoSe$_2$ the JT distortions are negligible so the doubly degenerate impurity levels remain degenerate and the exchange interactions do not change much with relaxation. 

\begin{figure}[t]
\includegraphics[width=7.6cm]{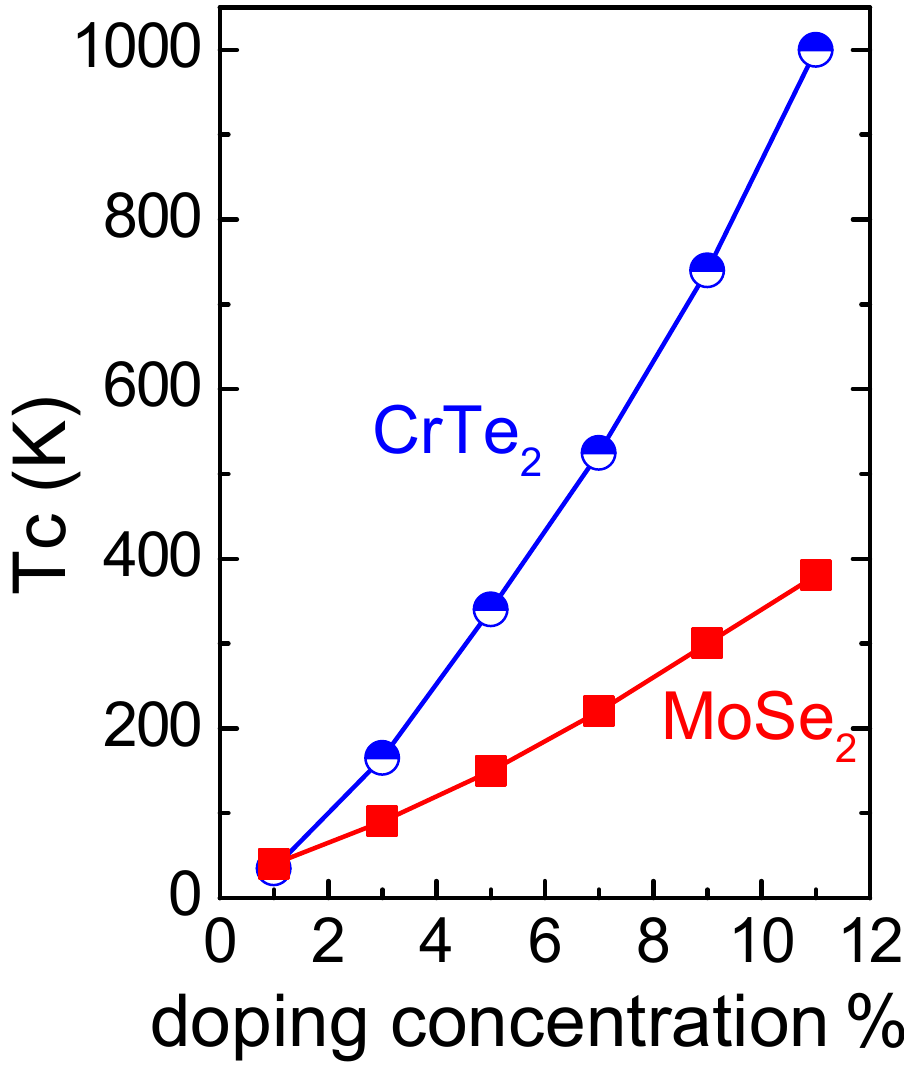}
\caption{\label{fig:Tc} Curie temperature as a function of doping concentration for V-doped MoSe$_2$ (red square) and CrTe$_2$ (blue half-filled circle) monolayers. 
}
\end{figure}

According to the results shown in \Cref{fig:exchange}, V-doped CrTe$_2$ and MoSe$_2$ exhibit the longest range exchange interactions of all the systems we have studied. We use the same Monte Carlo methodology as used for MoS$_2$:V in \cite{Gao:prb19a, *Gao:prb19b} to estimate their Curie temperatures ($T_{\rm C}$) and plot these as a function of the vanadium concentration $x$ in \Cref{fig:Tc}(b). Room temperature values of $T_{\rm C}$ are predicted for MoSe$_2$ and CrTe$_2$ monolayers for V dopant concentrations of 9\% and 5\%, respectively. In view of the comparable magnitude of the exchange interaction between very close V dopant pairs for CrTe$_2$ and MoSe$_2$, the longer range of the exchange interaction in CrTe$_2$ is pivotal in achieving RT ferromagnetism for low dopant concentrations.  

% ========== Figure 12 ==========
% ====================

%%%%%%%%10%%%%%%%%20%%%%%%%%30%%%%%%%%40%%%%%%%%50%%%%%%%%60%%%%%%%%70%%%%%%%%80
\section{Discussion}
\label{sec:disc}

%{\color{blue} Structure of the discussion section: paragraph 1-3, discuss my results and related previous work (p1:spin polarization; p2: SIA of single acceptor; p3:SIA of double acceptor);  paragraph 4. compared with recent experiments RT FM; paragraph 5, outlook for RT FM, itinerant ferromagnetism; paragraph 6, potential candidate: 1H CrTe2 }

We have explored how the magnetization of hole-doped MX$_2$ monolayers depends on the interplay of the exchange interaction of pairs of impurities with the exchange splitting \cite{footnote3}, JT distortion and magnetic anisotropy of single impurities. The fundamental conundrum in the quest for a high temperature magnetic semiconductor is the incompatibility of a long range exchange interaction mediated by effective-mass-like states and of robust local magnetic moment formation determined by strong on-site Hund's rule coupling. 
Before considering the experimental realisation of room-temperature ferromagnetic semiconductors by hole-doping MX$_2$ layers substitutionally, we use our computational results and analysis to review a number of recent computational studies based upon the same DFT plus supercell methodology we have also used. 

%%%%%%%%10%%%%%%%%20%%%%%%%%30%%%%%%%%40--------50--------60--------70--------80
\subsection{Other computational work}

 DFT is considered to be a suitable framework to use to study itinerant magnetism, both formally and in terms of practical calculations. 
 It correctly predicts Fe, Co and Ni to be the only transition metals that are ferromagnetic \cite{Gunnarsson:jpf76, Janak:prb77, Andersen:physbc77a} and has been used to study a wide range of properties of very diverse magnetic materials \cite{Kubler:00, Mohn:03}. 
 The detailed results of practical DFT calculations depend on which of a large number of exchange-correlation functionals is used. 
 Otherwise, different implementations of DFT can be used to reproducibly determine total energies and derived quantities if sufficient care is taken \cite{Lejaeghere:sc16}.
 Before comparing our results to those of other computational studies, it will be useful to review a number of factors that need to be considered when carrying out DFT calculations as well as a number of open issues. 
 
 All of the computational studies of doped MX$_2$ systems of which we are aware were performed by modelling substitutional dopants in $L \times M \times N$ repeated supercells of the type discussed in \Cref{sec:CM} with $L$ chosen to be unity to model a TMD monolayer and with a large thickness of vacuum inserted to minimize the interaction of the monolayer with its periodic image in the $c$ direction, perpendicular to the plane of the TMD monolayer. 
 In the plane of the monolayer, the size of the $M \times N$ lateral supercell determines the lowest concentration of dopant that can be modelled. 
 By looking at a single substitutional donor or acceptor dopant in an $N \times N$ supercell as a function of $N$ \cite{Gao:prb19b}, we can examine how closely we can approach the single impurity limit described by the effective mass theory (EMT) referred to in \Cref{sssec:BE}. 
 We already discussed in \Cref{sec:CM} how the near-universal choice of a GGA exchange-correlation functional leads to problems with the separation $\Delta_{K\Gamma}=\varepsilon_\upsilon(K)-\varepsilon_\upsilon(\Gamma)$ of the valence band maxima at the $\Gamma$ and $K/K'$ points of high symmetry compared to experiment and how this led us to instead use the L(S)DA approximation that describes $\Delta_{K\Gamma}$ much better. 
 
 The single impurity limit is important because (i) it is desirable to achieve room-temperature ferromagnetic ordering with as low a dopant concentration as possible in order to retain the semiconducting properties of the MX$_2$ host and (ii) single impurities can be studied experimentally to verify the calculations in a well-defined limit \cite{Lannoo:81, Bourgoin:83}. 
 As discussed at length for MoS$_2$:V(Nb and Ta) in \cite{Gao:prb19b}, many of the different results found in the literature can be understood in terms of the choice of exchange-correlation potential or some technical aspect of the calculations such as plane-wave cutoff, k-point sampling, vacuum thickness etc. We will not repeat this discussion here but instead consider two open issues that were not addressed in detail in \cite{Gao:prb19b}.

%-------10--------20--------30--------40--------50--------60--------70--------80 
\subsubsection{Self-interaction, LDA+U} 
\label{sssec:LDAU}
 
The first concerns how a single unpaired spin is treated in DFT.  
Consider the electronic structure of the JT-distorted MoTe$_2$:V impurity shown in \Cref{fig:PES}(c). 
In the absence of spin polarization, the Fermi level is seen to lie in the spin-degenerate $d_{xy}$ state corresponding to half a spin-up hole and half a spin-down hole. 
This unphysical feature (of fractions of indivisible particles) of DFT-LDA can be resolved by using spin-polarized DFT (SDFT) and a spin version of the LDA (LSDA) based upon the spin-polarized electron gas \cite{vonBarth:jpc72, Gunnarsson:prb76}. 
In the DFT for the total energy of a hydrogen atom \cite{Gunnarsson:prb74}, the Coulomb interaction of the single electron with itself (self-interaction) should be exactly cancelled by the exchange-correlation energy term, as it is in orbital-based Hartree-Fock. 
In the LDA the cancellation is very incomplete, leading to a total energy of about -0.9 Rydberg. 
The LSDA introduces a Hund's rule like interaction of the electron density leading to an electron with a single spin (up or down) and a total energy of about -0.97 Rydberg, much closer to the correct result. 

% ========== Figure 13 ==========
\begin{figure}[t]
\includegraphics[width=8.6cm]{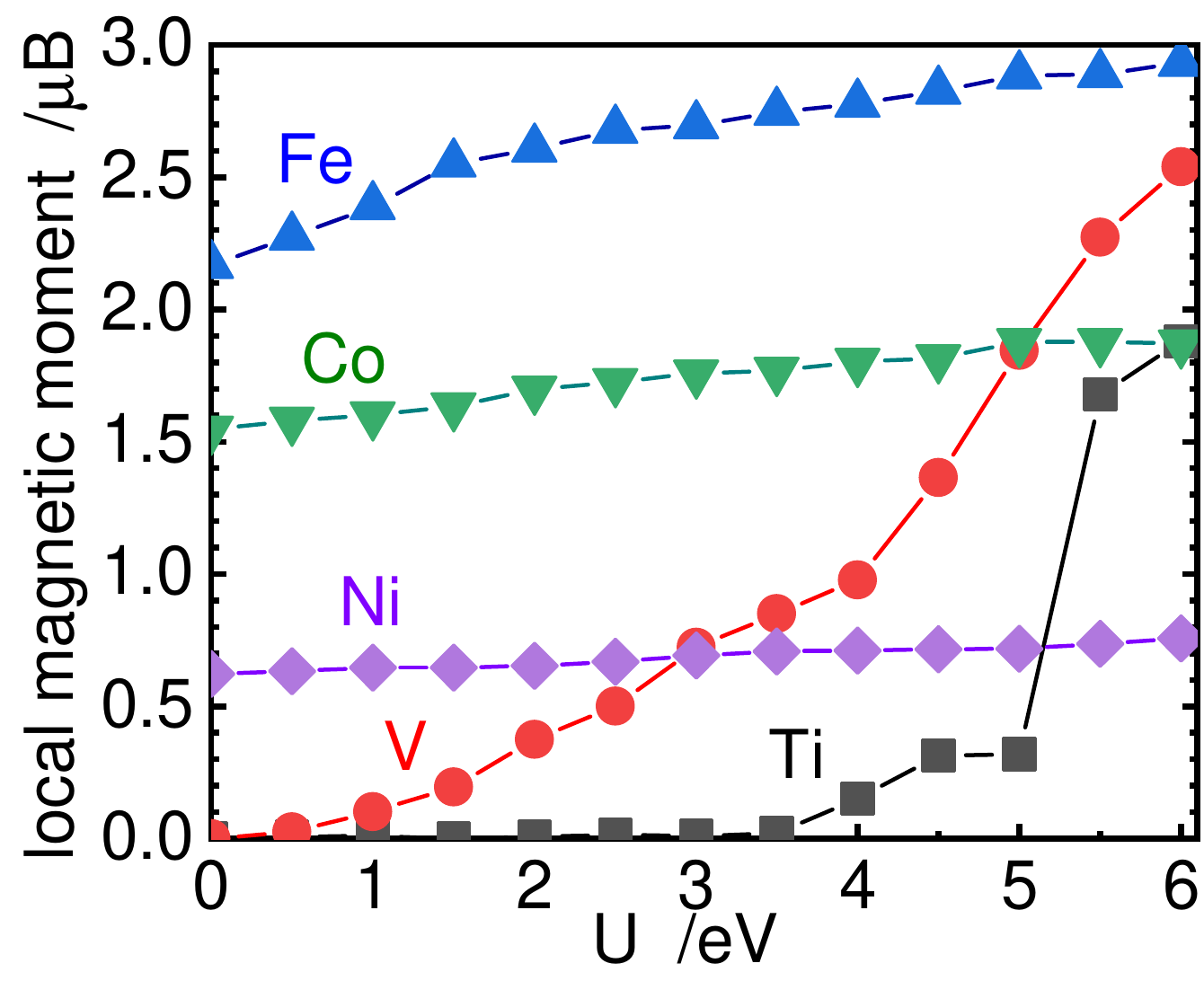}
\caption{\label{fig:LDA+U} Local magnetic moment per atom in $\mu_{\rm B}$ as a function of Coulomb $U$ (eV) on the 3$d$ orbitals for hcp Ti, bcc V, bcc Fe, hcp Co, fcc Ni.  
}
\end{figure}
% ====================

For many materials with partly filled and quite localized $d$ or $f$ shells, the LSDA ``Hund's rule'' solution was found to describe spectroscopic properties like the bandgap quite inadequately and attention focussed on the atomic description of these states in terms of Coulomb integrals like the Hubbard $U$ parameter \cite{Anisimov:prb91b, Solovyev:prb94, Liechtenstein:prb95} leading to an improved, mean-field description of the spectroscopic properties of such ``correlated'' materials termed L(S)DA+$U$ \cite{Dudarev:prb98}. A major disadvantage of the LDA+$U$ approach is that it is not known what value of $U$ should be used and it is usually chosen to reproduce some experimentally known quantity like the bandgap whereby the predictive capacity of DFT is lost. For example, we show in \Cref{fig:LDA+U} the value of the magnetic moment ``predicted'' by LSDA+$U$ as a function of the parameter $U$ (actually $U-J$ in \cite{Dudarev:prb98}) for a number of 3$d$ transition metals where we have used the experimental lattice constants and crystal structures. For bcc Fe, hcp Co and fcc Ni, the spin moment is overestimated for values of $U\sim 3-5\,$eV typically used while hcp Ti and bcc V are incorrectly predicted to be ferromagnetic. 

For single electron systems like the H atom, or single hole systems \cite{Bethe:86} like the single acceptors considered in this manuscript, there is no physical interaction of the electron or hole with itself, no question of the MX$_2$:V system being ``strongly correlated''. A justification for including a finite $U$ on the V impurity might be that it mimics a self-interaction correction. However, this is not the justification given in the literature.
  
%-------10--------20--------30--------40--------50--------60--------70--------80
\subsubsection{SP+SOC versus SOC+SP} 
\label{sssec:SPSOC}

% ========== Figure 14 ==========
\begin{figure*}[t]
\includegraphics[width=17.6cm]{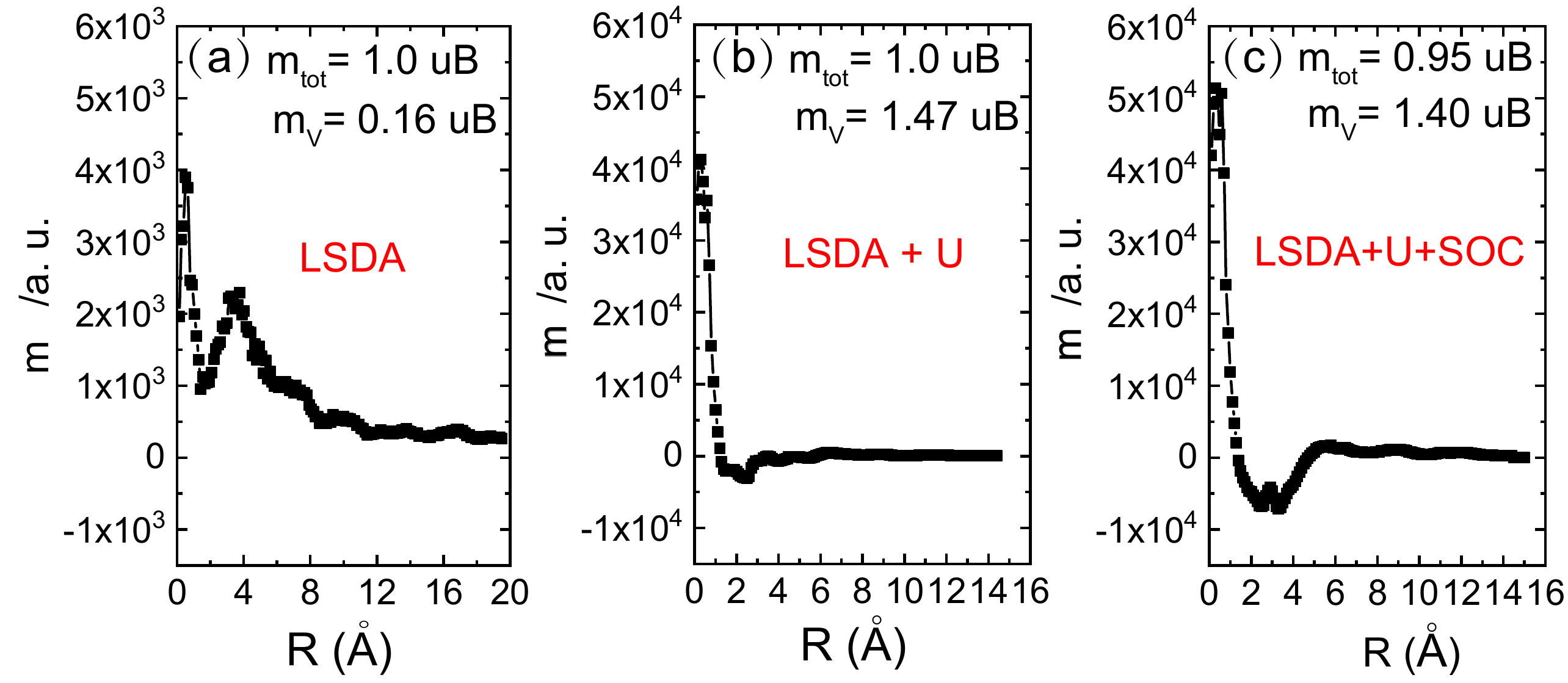}
\caption{\label{fig:mofR} Radial distribution of the magnetization arising from a single substitutional V acceptor in WSe$_2$ calculated for  
	(a) the LSDA using a 12$\times$12 supercell and using a 9$\times$9 supercell
	(b) the LSDA+$U$ with $U=3\,$eV, 
	(c) the LSDA+$U$ with $U=3\,$eV with SOC included in the SCF cycle.
}
\end{figure*}
% ====================

The second open issue we consider relates to the simultaneous inclusion of spin-polarization (SP) and spin-orbit coupling (SOC) or the order in which they should be treated if included as perturbations.  
To predict an ordering temperature $T_{\rm C}$, we (i) calculated the SIA in \Cref{sssec:SIA} by determining a spin-polarized solution of the Kohn-Sham equations and then included SOC as a perturbation (``force theorem'') and (ii) estimated the interatomic exchange interaction $J(R)$ in \Cref{sec:ResPI} by calculating the total energy of a pair of dopant atoms separated a distance $R$ with the on-site spins aligned (FM) and antialigned (AFM), without SOC; by interpreting the energy difference $E_{\rm FM}-E_{\rm AFM}$ in terms of a Heisenberg model, we arrived at  $J(R)$. 
Because of the large Bohr radius of the MX$_2$:V single acceptor $e'$ states (\Cref{TableVI}), most of the corresponding hole is not on the V site and the SOC for these states is largely determined by that of the host. 
For the MoS$_2$ host considered in \cite{Gao:prb19a, *Gao:prb19b}, the dominant exchange splitting of the $a'_1$ state was comparable in magnitude to the spin-orbit splitting of the $e'$ level and we did not attempt to include SOC in iterating the total energy to determine $E_{\rm FM}$ and $E_{\rm AFM}$. For compounds containing W and Se or Te, this argument is less likely to be valid; spin-polarization should ideally be included as a perturbation on top of SOC. The problem is that the spin-dependence of the spin-dependent LSDA exchange-correlation potential depends (sensitively) on the spin density and can only be determined by an iterative self-consistent field (SCF) procedure. Including SOC doubles the size of the Hamiltonian that must be diagonalized and becomes roughly eight times more expensive; a factor of two is regained because spin-up and spin-down no longer need to be treated separately. The resulting factor of four is probably an underestimate of the true computational cost because the small size of the magnetic moments and exchange-splitting makes many SCF iterations necessary. Including a finite value of $U$ in this procedure makes the magnetism more robust and we will examine the result of doing so below.    

%{\color{red}This  }
%As an example, consider MoSe$_2$:V with a total moment of $1\mu_{\rm B}$. Without $U$, only  $0.2\mu_{\rm B}$ of this moment is on the V impurity, the rest is on the surrounding Se and Mo atoms, parallel to that of the dopant. Including $U=3\,$eV on the V impurity increases the local moment of the dopant from $0.2\mu_{\rm B}$ to 0.67 $\mu_{\rm B}$ but at the cost of inducing moments on the nearest Se ligands that are antiparallel to that of the dopant while the total net magnetic moment is still 1 $\mu_{\rm B}$. This is important because it will affect the exchange interaction and how it depends on SOC. Without $U$, SOC reduces the local magnetic moment of the dopant by around 10 percent and the exchange interaction only decreases slightly. With $U$, as in the case of WSe$_2$:V, {\color{red}It would be a lot stronger if you could say what the effect of $U$ and SOC is on the coupling and the moments for one and the same system; jumping from MoSe$_2$ to  WSe$_2$ is weak as is ``greatly decrease the exchange interaction''.} SOC does not change the moment of the dopant much but enhances the moments of the Se closest to the dopant that are antiparallel to the dopant, which greatly decrease the exchange interaction. So how the SOC affect the magnetic moment depend on the Coulomb $U$ used in the calculations. 
 
%-------10--------20--------30--------40--------50--------60--------70--------80  
\subsubsection{Dopant-induced magnetic moments} 
\label{sssec:DIMM}

Inducing magnetism in MX$_2$ monolayers by doping them $p$-type with group IV and V transition metals (rather than with magnetic elements like Mn, Fe, Co, and Ni) has attracted a considerable amount of attention over the past decade. 
However, depending on the computational details, magnetic moments ranging from 0 \cite{Coelho:aem19, Dolui:prb13} to 2.4$\, \mu_{\rm B}$ \cite{Andriotis:prb14} are reported for single acceptors \cite{Yue:pla13, Lu:nrl14, Singh:acsami16, Li:cm16, Wu:pla18, Duong:apl19, Zhang:advs20, Pan:jpd20, Tiwari:tdma21}; and from 0 \cite{Singh:acsami16, Tiwari:tdma21, Dolui:prb13, Zhao:jac16} to 1.6$\, \mu_{\rm B}$ \cite{Andriotis:prb14} for double acceptors.
Because the width of the impurity band depends on the size of supercell used in a DFT calculation, we expect the magnetic moment to decrease as the supercell size is reduced causing the impurity bands to broaden \cite{Gao:prb19a, *Gao:prb19b}. 
We attribute the finding of single acceptors with magnetic moments per impurity of 0$\, \mu_{\rm B}$ \cite{Pan:jpd20, Dolui:prb13} or less than 1$\, \mu_{\rm B}$ \cite{Li:cm16, Wu:pla18} to the use of small supercells. 
Calculated magnetic moments show a strong dependence on the inclusion of a Hubbard $U$ $ \sim \,$ 3-6 eV on the $d$ orbitals of the dopant. 
This increases the splitting of the spin-up and spin-down dopant levels resulting in magnetic moments as large as 1.7$\, \mu_{\rm B}$ for single acceptors \cite{Andriotis:prb14, Singh:acsami16, Wu:pla18, Duong:apl19, Tiwari:tdma21}. 
In the context of our remarks in \Cref{sssec:LDAU}, we note that a nonmagnetic system can be forced to become spin-polarized if $U$ is larger than the separation of occupied and unoccupied levels. 
For example, with $U=0$, MoS$_2$:Ti was reported to be nonmagnetic \cite{Andriotis:prb14} but to become magnetic with an atomic moment of 1.6$\, \mu_{\rm B}$ with $U = 5.5 \,$eV. 
This large value of $U$ causes the $a'_1$ state of MoS$_2$:Ti (see \Cref{fig:impuritylevel}) to split so that the hole is transferred to the $e'$ state yielding a magnetic moment of well over 1 $\mu_{\rm B}$ \cite{Andriotis:prb14}. 
However, even with $U = 4 -6\,$ eV, Tiwari did not find magnetism for Ti-doped MoS$_2$; presumably because of some other computational choice such as an inadequate supercell size \cite{Tiwari:tdma21}. 

The spatial distribution of the magnetization is strongly affected by the inclusion of a Hubbard $U$.  
This is illustrated in \Cref{fig:mofR} for WSe$_2$:V where we find a total magnetic moment of 1.0$\, \mu_{\rm B}$ both in the LSDA and in the LSDA+$U$ with $U=3\,$eV. It is only slightly reduced to   0.95$\, \mu_{\rm B}$ when SOC is included simultaneously with a $U=3\,$eV.
However, where the moment on the V impurity is only 0.16$\, \mu_{\rm B}$ in the LSDA, including a Hubbard $U=3\,$eV leads to a much larger V moment of 1.47$\, \mu_{\rm B}$ (1.40$\, \mu_{\rm B}$ when SOC is included); this is compensated by antiparallel moments on the neighbouring W and Se.
This difference in spatial distribution is in principle amenable to experimental study using spin-polarized scanning tunneling microscopy \cite{Wiesendanger:rmp09}.
We will see the consequences of this for the magnetic coupling between pairs of V dopants below.

Although the Hubbard $U$ included in the calculations can account for the correlation effect of $d$ electrons as on Mn \cite{Gil:jpcm14}, the choice of numerical value is notoriously fraught.  
For a (two-electron or) two-hole system, the Coulomb repulsion represented by a Hubbard $U$ is more easily justified than for a (single electron or) hole especially when the Bohr radius is as large as we have found; in the single particle case the use of a Hubbard $U$ might be justified in terms of self-interaction correction but that comes at the expense of predictive capability. 

%-------10--------20--------30--------40--------50--------60--------70--------80
\subsubsection{Comparison with the work of Duong et al.}
\label{sssec:Duong}

Though many TMD host materials contain heavy elements for which the SOC is quite large, not much attention has been paid to the effect of this SOC in previous work. 
An interesting exception is a computational study of WSe$_2$:V by Duong {\it et al.}  \cite{Duong:apl19} who found that including SOC led to an increase of the spatial extent of the magnetization of a single substitutional V impurity and proposed this as an explanation for observations of long range ferromagnetic ordering in doped MX$_2$ systems at elevated temperatures \cite{Pham:am20, Guguchia:sca18, Coelho:aem19, Yang:aem19, Hu:acsami19, Zhang:advs20, Yun:advs20}. 
Using {\sc vasp}, we have repeated these calculations, iterating the calculations to selfconsistency including SOC with a Hubbard $U=3\,$eV on the V impurity. 
We confirm that simultaneous inclusion of SOC and spin-polarization leads to an enhancement of the (small) magnetic moments on W atoms far from the dopant. However, we also find that this occurs at the expense of larger antiparallel magnetic moments on the nearest neighbour W and Se atoms, compare \Cref{fig:mofR}(b) and \Cref{fig:mofR}(c).  

% ========== Figure 15 ==========
\begin{figure}[t]
\includegraphics[width=8.6cm]{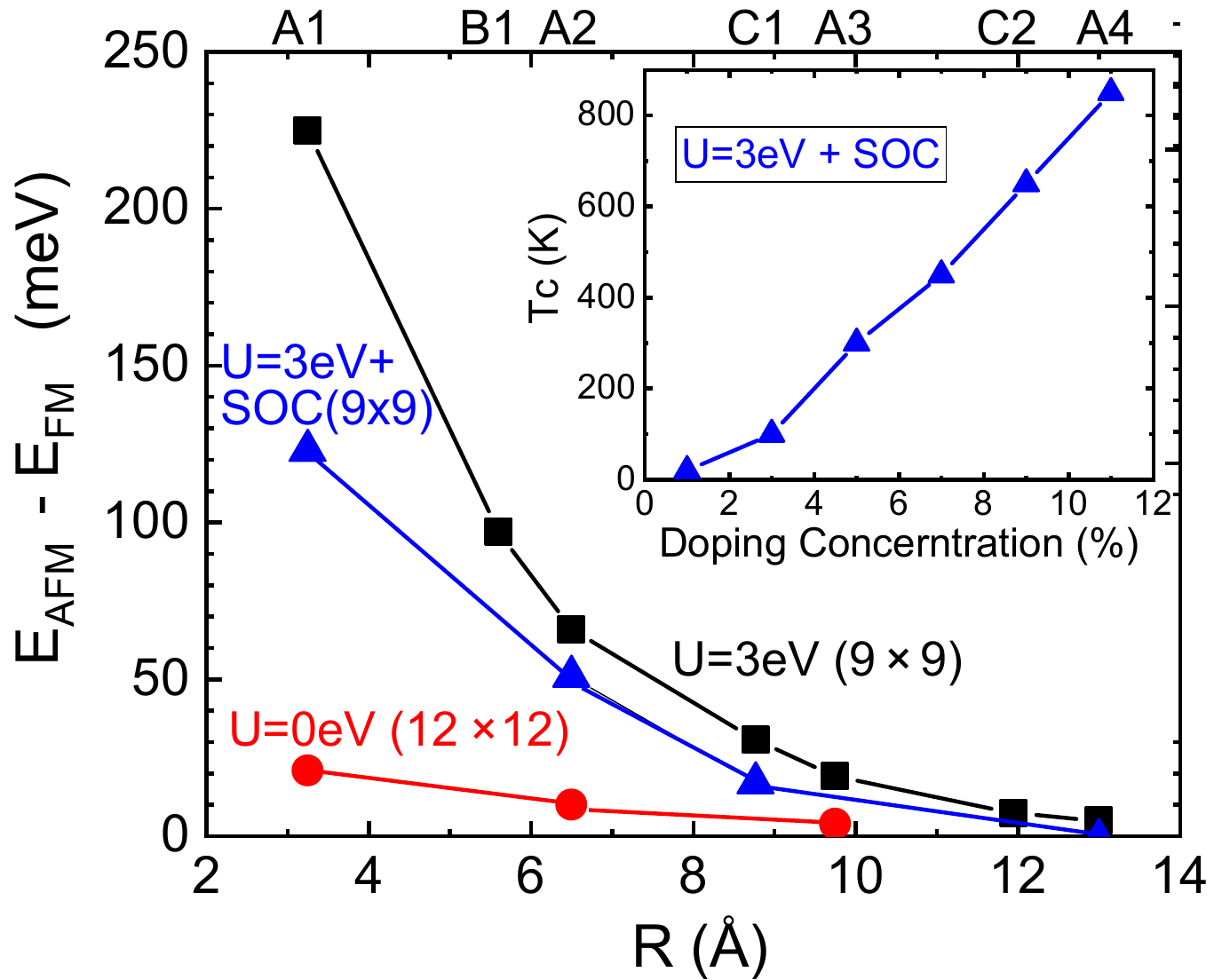}
\caption{\label{Fig15} Difference between the total energies of antiferromagnetic and ferromagnetic states for pairs of substitutional V impurities including relaxation in 9$\times$9 supercells plotted as a function of their separation for WSe$_2$ monolayers when $U=3\,$eV is included on V. The LSDA results calculated in a 12$\times$12 supercell are included for comparison ($U=0\,$eV). The solid lines are a guide to the eye. The labelling of the W sites surrounding the central impurity atom used on the top $x$ axis is the same as that used in Fig.3 of \cite{Gao:prb19b}. 
}
\end{figure}
% ====================

To see the effect this has on the exchange coupling, we performed total energy calculations for parallel and antiparallel aligned V pairs at a number of separations (using a 9$\times$9 supercell because of the computational expense when the calculations are iterated with SOC).
The upper curve (black squares) in \Cref{Fig15} shows this coupling energy when SOC is not included. 
Compared to the LSDA results presented in the previous section, we see a huge increase in the strength of the coupling energy, from a maximum value of $\sim 35\,$meV for V impurities on neighbouring Mo sites (\Cref{fig:exchange}) to 225 meV (\Cref{Fig15}).
This can be traced to the effect of $U$ increasing the local magnetic moment of V from 0.16 $\mu_B$ to 1.47 $\mu_B$, where the antiparallel magnetic moments on the W and Se neighbours limit the total magnetic moment to 1 $\mu_B$.
When SOC is included, the exchange interaction is reduced substantially, from 225 to 123 meV (blue triangles in \Cref{Fig15}) but remains much larger than the LSDA value; the results are dominated by the use of a Hubbard $U$.  
When the impurities are further apart (configuration A4 in Fig.3 of \cite{Gao:prb19b}), SOC reduces the corresponding energy difference from 5.4 to -1.4 meV. 
This can be attributed to the SOC enhancement of the antiparallel magnetic moment on the W atoms closest to the V from -0.02 $\mu_{\rm B}$/W to -0.06 $\mu_{\rm B}$/W, which ``screen'' the exchange interaction between V pairs. 
The effect of $U$ is to predict much higher critical temperature $T_{\rm C}(x)$ for high dopant concentrations $x$; see the inset in \Cref{Fig15}. 

Including a Hubbard $U$ enhances the moment on the impurity site and leads to much larger exchange interactions. 
It also makes the iteration of the Kohn-Sham equations of DFT much more stable when  SOC is included.
The effect of SOC is large for WSe$_2$:V because of the large SOC of the 5$d$ electrons of W. 
Tests for MoSe$_2$:V, indicate that the effect of SOC on the exchange interactions shown in \Cref{fig:exchange} is much smaller so that our estimates of the ordering temperature should be a lower limit. 
While choosing intermediate values of $U$ will lead to intermediate $T_{\rm C}(x)$ curves, the basic underlying model of exchange coupling between effective mass like impurity states remains the same. 
 
%-------10--------20--------30--------40--------50--------60--------70--------80
\subsubsection{Comparison with the work of Tiwari et al.}
\label{sssec:Tiwari}

In their calculations including SOC, Tiwari {\it et al.} have gone a step further than Duong {\it et al.} \cite{Duong:apl19} by calculating the anisotropic interatomic exchange coupling $J_{ij}(R)$ and SIA with DFT calculations (including a Hubbard $U$ ranging from 4-6~eV) and combining these results in parameterized form with Monte-Carlo simulations to estimate ordering temperatures for a large number of dopants in TMD monolayers \cite{Tiwari:tdma21}. 
Of relevance for the present discussion are their results for substitutional V and Ti in MoS$_2$, MoSe$_2$, MoTe$_2$, WS$_2$ and WSe$_2$. 
For all MX$_2$:Ti systems, they found negligible magnetic moments, which agrees with our findings for the double acceptor doped diselenides and ditellurides when the JT distortion is included, \Cref{TableIII}. 

For MX$_2$:V, Tiwari {\it et al.} found FM ordering for the disulfides and diselenides but not for MoTe$_2$:V (which has a large magnetic moment of V up to 2 $\mu_{\rm B}$ as listed in his Supplementary Table. 3 but orders antiferromagnetically). 
For their most promising system, MoSe$_2$:V, Tiwari found $J(R)$ has a maximum value of $\sim 10\,$meV comparable to the value of $\sim$ 25 meV we report in \Cref{fig:exchange}(b) where $J(R)$ is half of the total energy difference between AFM and FM. 
However, they reported an ordering temperature of 224~K with a concentration of 15\% V compared to our finding of ordering with this temperature for half of this concentration. 
We can trace the difference in this case to the choice of Hamiltonian used in the Monte Carlo simulations. 
For the doped systems, Tiwari {\it et al.} used a parameterized Heisenberg Hamiltonian that includes anisotropic exchange (less than 2\%) and a single ion anisotropy of less than 1 meV (\cite{Tiwari:tdma21}, Supplementary Material). The Heisenberg Hamiltonian allows the spin to rotate in three dimensions at finite temperature. In our work, the large single-ion anisotropy (as high as 10 meV per dopant) allows us to approximate the doped system with an Ising model. 
We attribute the difference in ordering temperatures to Tiwari's much smaller anisotropy. The differences in the single ion anisotropy can be understood in terms of the technical details of the two sets of calculations. Because he uses the GGA, we expect Tiwari's ordering of the $a'_1$ and $e'$ defect states to be incorrect in the single impurity limit because the GGA describes the relative positions of the $\Gamma$ and $K/K'$ VBM poorly compared to LDA. Using a Hubbard $U$ to ``account for the electron-electron interaction'' of the single hole greatly enhances the exchange splitting of $a'_1$ and $e'$ levels. We reproduced his calculations and found the holes reside mainly in the $a'_1$ levels that do not contribute to the magnetic anisotropy energy. Tiwari's calculation of an anisotropic interatomic exchange coupling $J_{ij}(R)$ based upon inclusion of SOC in the self-consistent SP iterations is in principle better than out treatment but comes at the price of limiting the size of supercell that he could use; 7$\times$7 in his case versus our 12$\times$12. 

%%%%%%%%10%%%%%%%%20%%%%%%%%30%%%%%%%%40--------50--------60--------70--------80
\subsection{Experiments}

%==========TableVIII=========
\begin{table}[b]
\caption{Single hole dopants in MX$_2$ TMDs for which room temperature magnetic ordering has been reported. }
\begin{ruledtabular}
\begin{tabular}{llll}
M \verb+\+ X & S              & Se & Te \\
\hline
Cr    & -- & -- & -- \\
Mo    & V \cite{Hu:acsami19}  & -- & V \cite{Coelho:aem19}; Ta \cite{Yang:aem19} \\
W     & V \cite{Zhang:advs20} & V \cite{Yun:advs20, Pham:am20, Yun:am22} & --  \\
\end{tabular}
\end{ruledtabular}
\label{TableVIII}
\end{table}
%==========TableVIII=========

Long-range defect- and impurity-induced magnetic ordering has been reported for the TMDs shown in \Cref{TableVIII},
%\cite{Guguchia:sca18, Coelho:aem19, Yang:aem19, Hu:acsami19, Pham:am20, Zhang:advs20, Yun:advs20} 
with room temperature ordering reported for the monovalent hole dopants, mostly vanadium, for dopant concentrations as low as 0.1\% \cite{Yun:advs20}, 0.2\% \cite{Coelho:aem19} or 0.4\% \cite{Zhang:advs20}. 
Optimum concentrations have yet to be determined and the origin of the high temperature ferromagnetism at these low doping concentrations is still an open question. 
Our calculations support the qualitative finding of ferromagnetism and provide a microscopic picture to interpret experiments but do not explain why the coupling is so strong at such low concentrations. 
It is conceivable that this is just a quantitative failing of the LDA; for the moment we focus on qualitative trends to suggest directions for further research. 

The sensitivity of the magnetic properties of a monolayer of WSe$_2$:V to a back-gate bias has been invoked to argue for the relevance of an RKKY model \cite{Yun:advs20}. 
In the RKKY model \cite{Kubler:00, Mohn:03}, localized atomic magnetic moments (transition metal $d$ electrons; rare-earth $f$ electrons) couple via the free electron gas between them (transition metal $s$ electrons; rare-earth $sd$ electrons). 
A gate voltage that can modulate the density of the electron gas changes the coupling between the localized magnetic moments. 
However, this is not the physical picture presented by our calculations nor indeed by those of Duong {\it et al.} \cite{Duong:apl19}. 
Our DFT calculations show that there are no localized atomic moments nor is there a free-electron-like gas. 
In the single impurity limit, there are single hole states bound to the valence band maxima with host M $d$ character that are spread over many host M sites as well as the dopant atom. 
As the dopant concentration is increased, these single spins couple ferromagnetically and the impurity states form a narrow, partly-filled, impurity band whose occupancy and the coupling between the spins, will be modified by gating.
The manner in which it is changed should in principle be accessible to DFT calculations but this goes well beyond the scope of the present manuscript.  

The occurrence of an optimum doping concentration of order 2\% in WS$_2$:V by Zhang {\it et al.} \cite{Zhang:advs20} is indirect support for our finding that the ferromagnetic coupling between two vanadium impurity atoms in WS$_2$ is ``quenched'' below a critical separation \cite{Gao:prb19a, *Gao:prb19b} which we explained in terms of the preference for vanadium pairs to form singlet states in TM disulphides as a consequence of the near-degeneracy of the impurity $a'_1$ and $e'$ states when they are close. However, in TM diselenides and ditellurides, the $a'_1$ levels lie in the valence band and the quenching assumed by Pham {\it et al.} \cite{Pham:am20} for WSe$_2$:V is not supported by our LDA calculations.     

In their calculations for WSe$_2$:V, Duong {\it et al.} \cite{Duong:apl19} find a small magnetic moment on the W nearest-neighbour sites of the central impurity V that is antiparallel to the V moment as well as the moments on the remaining W hosts. 
We confirm this quantum mechanical description of the spatial distribution of a single hole state and note that it involves a single spin degree of freedom. Applying an external magnetic field or changing the temperature will certainly change this spatial distribution. However, the picture invoked by Pham {\it et al.} \cite{Pham:am20} to explain an observed {\it increase} of magnetization with temperature implies that the moments on the central impurity and on the neighbouring W atoms are independent degrees of freedom and this explanation cannot be correct in detail. 

%Yun enhanced the magnetizations through high temperature heat-treatment that creates more Se vacancies near V in WSe$_2$ \cite{Yun:am22}. {\color{red}Do we really want to discuss this?}

In view of the key role that they play in the ferromagnetism of hole-doped MX$_2$ systems, it is important to verify the existence of the ($a'_1$ and) $e'$ impurity states bound to the VBM that we have identified in all MX$_2$ systems studied, \Cref{fig:impuritylevel}. 
One difficulty with finding these impurity levels is that experimental techniques like photoemission only detect occupied states; hole states can only be detected if they are occupied by e.g. gating or by using inverse photoemission spectroscopy. 
Using scanning tunnelling spectroscopy, Mallet observed bound impurity states at 140 meV above the VBM in a negatively charged V-doped WSe$_2$ monolayer whose Fermi level was determined by it having been grown on top of $n$-type epitaxial graphene \cite{Mallet:prl20}.
Occupation of the hole state with an electron will quench the Coulomb binding of the V impurity and should substantially reduce the binding energy of about 60 meV that we find for WSe$_2$:V, \Cref{fig:impuritylevel}. Instead, Mattel {\it et al.} interpret their data in terms of a V-related hole state with a binding energy of 140 meV. 
More detailed modelling and more extensive experiments are needed to resolve this discrepancy. 
Mallet's finding of three bound impurity states in a V$_2$ dimer does seem to imply the existence of multivalley impurity states. 

%%%%%%%%10%%%%%%%%20%%%%%%%%30%%%%%%%%40%%%%%%%%50%%%%%%%%60%%%%%%%%70%%%%%%%%80
\subsection{Outlook}

In spite of the large effective Bohr radius ($\sim$ 9.4 \AA) and SIA ($\sim$ 4.7 meV) that we find for single substitutional V impurities in WSe$_2$, our study of the exchange interaction between V impurities does not predict WSe$_2$:V monolayers to be the most favourable system for realizing room temperature ferromagnetism.
Instead, 1H-CrX$_2$:V, MoSe$_2$:V and MoTe$_2$:V are all predicted to exhibit stronger ferromagnetism than WSe$_2$:V (in the absence of a Hubbard $U$), \Cref{fig:exchange}. 
Edwards and Katsnelson have argued in the context of CaB$_6$ that the effective interaction and Curie temperature predicted by the Stoner criterion will not be reduced by correlation effects or spin wave excitations \cite{Edwards:jpcm06}. 
In this sense, systems with shallow and narrow impurity bands are very favourable for realizing high Curie temperature ferromagnetic semiconductors.
Our Monte-Carlo simulations identify 1H-CrTe$_2$ and MoSe$_2$ as the most promising of the systems we have considered.
In view of the experimental observation of room temperature ordering in WSe$_2$:V \cite{Yun:advs20, Pham:am20}, the outlook therefore appears very bright.

The failure for CrX$_2$ to adopt the semiconducting hexagonal phase presents a serious challenge \cite{Su:ami19}, in particular for realizing our prediction for CrTe$_2$:V. Because DFT studies for MX$_2$ monolayers indicate that the H structure is more stable than the T structure \cite{Ataca:jpcc12} and because of the large Bohr radius we find for all V doped systems, \Cref{TableVI}, it may  be simpler to start with MoSe$_2$ and explore the properties of the Cr$_x$Mo$_{1-x}$(Te$_y$Se$_{1-y})_2$ alloy system rather than trying to prepare V-doped CrTe$_2$ itself. 

%%%%%%%%10%%%%%%%%20%%%%%%%%30%%%%%%%%40%%%%%%%%50%%%%%%%%60%%%%%%%%70%%%%%%%%80
\section{Summary and Conclusions}
\label{sec:sumconc}

In this paper, a detailed study of hole doping of MoS$_2$ monolayers with single acceptors \cite{Gao:prb19a, *Gao:prb19b} has been extended to MX$_2$ monolayers with M = Cr, Mo, W and X = S, Se, Te that are doped with single (V, Nb, Ta) and double (Ti, Zr, Hf) acceptor dopants with a view to realizing a room temperature ferromagnetic semiconductor. We identified the three impurity bound states induced by the substitutional dopants of group IV and V transition metal in MX$_2$ monolayers and studied the competition between spin-polarization and JT distortion that determines the defect ground state in the single impurity limit. We included spin-orbit coupling in order to determine the magnetic anistropy in this limit, the so-called single ion anisotropy. This is out-of-plane for the single acceptors but  in-plane for the double acceptors which are therefore not suitable candidates for realizing magnetic semiconductors. Of the single acceptors, the LSDA exchange splitting is most robust for V making it the candidate of choice to study the exchange interaction between two impurity atoms. Intrinsic defects may enhance the local magnetic moment in $p$-doped MX$_2$ monolayers \cite{Hu:acsami19, Yun:am22} but will localize the holes and reduce the range of the interatomic exchange interaction. 

To estimate ordering temperatures, we calculated the energy difference between parallel (ferromagnetic) and antiparallel (antiferromagnetic) aligned pairs of V moments as a function of their separations in larger supercells and extracted from this energy difference an interatomic exchange interaction $J(R)$.   
Singlet states formed by close V pairs in MoS$_2$:V and WS$_2$:V monolayers represent a quenching of the magnetic moment (other single acceptors are similar to V). 
To avoid this quenching, the holes should occupy degenerate $e'$ levels as happens in single acceptor doped MSe$_2$ and MTe$_2$ monolayers; this also gives rise to large single ion anisotropies because of the large spin orbit interaction at the $K/K'$ valley of MX$_2$ monolayers from which the $e'$ levels derive. 
Partial occupation of these degenerate levels leads to Jahn-Teller distortions that can significantly reduce the exchange interaction as happens for MoTe$_2$:V, \Cref{fig:MoTe-exchange}. However, for WSe$_2$:V, MoSe$_2$:V, and CrX$_2$:V monolayers we find negligible JT distortions which can be attributed to the $e'$ levels being very extended as indicated by their effective Bohr radius. 
$J(R)$ is used in Monte-Carlo calculations to estimate the ordering temperature as a function of the dopant concentration and leads us to identify 1H CrTe$_2$:V and  MoSe$_2$:V as the most promising candidates to explore experimentally. The possibility of preparing M$_x$M$'_{1-x}$(X$_y$X$'_{1-y})_2$ alloys presents numerous  possibilities to tailor material properties.   

% ========== ACKNOWLEDGMENT ==========
\begin{acknowledgments}
This work was financially supported by the ``Nederlandse Organisatie voor Wetenschappelijk Onderzoek'' (NWO) through the research programme of the former ``Stichting voor Fundamenteel Onderzoek der Materie,'' (NWO-I, formerly FOM) and through the use of supercomputer facilities of NWO ``Exacte Wetenschappen'' (Physical Sciences). Y.G. thanks AHNU start-up grant for financial support.
\end{acknowledgments}
% ====================



\begin{thebibliography}{114}
\bibitem [{foo({\natexlab{a}})}]{footnote2}%
  \BibitemOpen
  \href@noop {} {}\bibinfo {note} {{Even if the impurity
  bandwidth is much smaller than the exchange splitting, incompletely filled
  states can be degenerate at the Fermi level with noninteger occupation
  numbers \cite{Janak:prb78, Jones:rmp89}}}\BibitemShut {NoStop}%
\bibitem [{foo({\natexlab{b}})}]{footnote1}%
  \BibitemOpen
  \href@noop {} {}\bibinfo {note} {{In the simplest case,
  $x$ is the symmetry-lowering displacement of a single atom. More generally,
  it represents a ``collective coordinate'' describing some more complex
  distortion of the central and surrounding atoms.}}\BibitemShut {Stop}%
\bibitem [{foo({\natexlab{c}})}]{footnote3}%
  \BibitemOpen
  \href@noop {} {}\bibinfo {note} {{A single electron or
  hole does not experience a net exchange interaction. However, in the
  extension of density functional theory to spin, the introduction of a
  spin-dependent exchange-correlation potential \cite{vonBarth:jpc72,
  Gunnarsson:prb76} leads to an exchange splitting of the Kohn-Sham eigenvalues
  even for a hydrogen atom \cite{Gunnarsson:prb74} and a lowering of the total
  ground state energy. The scheme shown in \Cref{fig:JT} represents these
  Kohn-Sham eigenvalues.}}\BibitemShut {Stop}%
\end{thebibliography}
\end{document}